\documentclass[5p, review]{elsarticle}
\pdfoutput=1
\usepackage{booktabs} 
\usepackage{tabularx}
\usepackage{multirow}
\usepackage{makecell}
\usepackage[utf8]{inputenc}
\usepackage{makecell}
\usepackage{subcaption}
\usepackage{xcolor}
\usepackage{pgfplots}
\usepackage{tikz}
\usetikzlibrary[patterns]
\usepackage{setspace}
\usepackage{hyperref}
\usepackage{gensymb}
\usepackage{xspace}
\usepackage{url}

\usepackage{algorithm}
\usepackage{algorithmic}

\usepackage{amsmath}
\usepackage{amssymb}

\DeclareMathOperator*{\argmax}{argmax} 

\newcommand{\approach}{\textsc{OSM2KG}\xspace}

\makeatletter
\def\ps@pprintTitle{%
 \def\@oddfoot{Accepted manuscript, Future Generation Computer Systems, DOI: \url{https://doi.org/10.1016/j.future.2020.11.003}}}
\makeatother

\newdefinition{definition}{Definition}

\def\hyph{-\penalty0\hskip0pt\relax}

\begin{document}

\title{Linking OpenStreetMap with Knowledge Graphs - Link Discovery for Schema-Agnostic Volunteered Geographic Information}

\author{Nicolas Tempelmeier$^1$}
\ead{tempelmeier@L3S.de}

\author{Elena Demidova$^{1,2}$}
\ead{elena.demidova@cs.uni-bonn.de}

\tnotetext[t1]{\textcopyright 2020. This manuscript version is made available under the CC-BY-NC-ND 4.0 license http://creativecommons.org/
licenses/by-nc-nd/4.0/}

\address{$^{1}$L3S Research Center, Leibniz Universit\"at Hannover, Germany}
\address{$^{2}$Data Science \& Intelligent Systems Group (DSIS), University of Bonn, Germany}

\begin{abstract}
Representations of geographic entities 
captured in popular knowledge graphs such as Wikidata and DBpedia are often incomplete. 
OpenStreetMap (OSM) is a rich source of openly available, volunteered geographic information that has a high potential to complement these representations. 
However, identity links between the knowledge graph entities and OSM nodes are still rare. 
The problem of link discovery in these settings is particularly challenging due to the lack of a strict schema and heterogeneity of the user-defined node representations in OSM.
In this article, we propose \texttt{OSM2KG} - a novel link discovery approach
to predict identity links between OSM nodes and geographic entities in a knowledge graph.
The core of the \texttt{OSM2KG} approach is a novel latent, compact representation of OSM nodes 
that captures semantic node similarity in an embedding. 
\texttt{OSM2KG} adopts this latent representation to train a supervised model for link prediction and utilises existing links between OSM and knowledge graphs for training. 
Our experiments conducted on several OSM datasets, as well as the Wikidata and DBpedia knowledge graphs, demonstrate that \texttt{OSM2KG} can reliably discover identity links. 
\texttt{OSM2KG} achieves an F1 score of 92.05\% on Wikidata and of 94.17\% on DBpedia on average, which corresponds to a 21.82 percentage points increase in F1 score on Wikidata compared to the best performing baselines.
\end{abstract}

\maketitle

\section{Introduction}
\label{sec:introduction}

OpenStreetMap\footnote{OpenStreetMap is a trademark of the OpenStreetMap Foundation, and is used with their permission. We are not endorsed by or affiliated with the OpenStreetMap Foundation.} (OSM) has recently evolved as the key source of openly accessible volunteered geographic information (VGI) for many parts of the world, building a backbone for a wide range of real-world applications on the Web and beyond \cite{JokarArsanjani2015}. Prominent examples of OSM applications include mobility and transportation services such as route planners \cite{epub34045}, public transportation information sites\footnote{\href{http://www.\%C3\%B6pnvkarte.de}{\texttt{http://www.öpnvkarte.de}}} and Global Positioning System (GPS) tracking\footnote{\url{https://gitlab.com/eneiluj/phonetrack-oc}}, as well as geographic information services\footnote{\url{https://histosm.org/}} and spatial data mining.

The OSM data is produced by a large number of contributors (approx. 5.6 million in August 2019\footnote{Statistics from \url{https://www.openstreetmap.org/stats/data_stats.html}}) and lacks a pre-defined ontology. 
The description of geographic entities in OSM (so-called ``OSM nodes'') includes few mandatory properties such as an identifier and a location as well as a set of user-defined key-value pairs (so-called ``tags'').
As a result, the representations of OSM nodes are extremely heterogeneous.
The tags provided for the individual OSM nodes vary highly \cite{Touya2015}. 

Knowledge graphs (KGs), i.e., graph-based knowledge bases \cite{FarberBMR18}, including Wikidata \cite{Vrandecic:2014}, 
DBpedia \cite{LehmannIJJKMHMK15}, YAGO2 \cite{HOFFART2013} and EventKG \cite{GottschalkD19} are a rich source of semantic information for geographic entities, including for example cities and points of interest (POIs). 
This information, typically represented according to the RDF data model, 
has a high and so far, mostly unexploited potential for semantic enrichment of OSM nodes. 
An interlinking of OSM nodes and geographic entities in knowledge graphs can bring semantic, spatial, and contextual information to its full advantage and facilitate, e.g., spatial question answering \cite{Punjani:2018} and semantic trip recommendation \cite{www19:tourrec}.

Interlinking of OSM and knowledge graphs has recently attracted interest in the Wikidata\footnote{\url{https://www.wikidata.org/wiki/Wikidata:OpenStreetMap}} and OSM\footnote{\url{https://wiki.openstreetmap.org/wiki/Proposed\_features/Wikidata}} communities.
Our analysis results, presented in Section \ref{sec:motivation}, illustrate that the coverage of the existing interlinking between the OSM nodes and Wikidata entities varies significantly across entity types and geographic regions. For example, in a recent OSM snapshot of Germany (referred to as \texttt{OSM-DE}), cities are linked more often (73\%) than less popular entities like mountains (5\%). For another example, there are 42\% more linked OSM nodes in the OSM snapshot of Germany than in that of Italy (\texttt{OSM-IT}).
In practice, the interlinking of OSM nodes with semantic reference sources such as Wikidata or DBpedia is typically conducted manually by volunteers (and sometimes companies, see, e.g., \cite{GaneshBlog}). 

The problem of OSM link discovery is particularly challenging due to the heterogeneity of the OSM node representations. Other factors affecting the effectiveness of OSM node disambiguation in the context of link discovery include place name ambiguity and limited context \cite{Gritta2018}. 
Furthermore, geographic coordinates in the VGI sources such as OSM often represent the points of community consensus rather than being determined by objective criteria \cite{SLHA11} and can thus vary significantly across sources. 
For example, an average geographic distance between the coordinates of the corresponding entities in Germany in the OSM and Wikidata datasets is 2517 meters.
This example illustrates that geographic coordinates alone are insufficient to effectively discover identity links between the corresponding entities in VGI sources.

Although research efforts such as the LinkedGeoData project \cite{SLHA11} and Yago2Geo \cite{10.1007/978-3-030-30796-7_12} have been conducted to lift selected parts of OSM data in the Semantic Web infrastructure to facilitate link discovery, these efforts typically rely on manually defined schema mappings. Maintenance of such mappings does not appear feasible or sustainable, given the large scale, and openness of the OSM schema. 
Therefore, link discovery approaches that can address the inherent heterogeneity of OSM datasets are required.

In this article, we propose the novel \approach link discovery approach to establish identity links between the OSM nodes and equivalent geographic entities in a knowledge graph. 
\approach  addresses OSM's heterogeneity problem through a novel latent representation of OSM nodes inspired by the word embedding architectures \cite{NIPS2013_5021}. 
Whereas embeddings have recently gained popularity in several domains, their adoption to volunteered geographic information in OSM is mostly unexplored. 
In contrast to state-of-the-art approaches to link discovery in OSM (such as \cite{10.1007/978-3-030-30796-7_12, SLHA11}), 
\approach does not require any schema mappings between OSM and the reference knowledge graph.

The core of the \approach approach is a novel latent representation of OSM nodes 
that captures semantic node similarity in an embedding. 
\approach learns this latent, compact node representation automatically from OSM tags.
To the best of our knowledge \approach is the first approach to 
address the heterogeneity of the OSM data by a novel embedding representation.
This embedding representation is created in an unsupervised fashion 
and is task-independent. 
The embedding systematically exploits the co-occurrence patterns of the OSM's key-value pairs to capture their semantic similarity.  
Building upon this embedding, along with spatial and semantic information in the target knowledge graph, \approach builds a supervised machine learning model to predict missing identity links.
To train the proposed link prediction model, we exploit publicly available community-created links between OSM, Wikidata, and DBpedia as training data.

%

The key contribution of our work is the novel \approach link discovery approach to infer missing identity links between OSM nodes and geographic entities in knowledge graphs, including:
\begin{itemize}
\item A novel unsupervised embedding approach to infer latent, compact representations that capture 
semantic similarity of heterogeneous OSM nodes.
\item A supervised classification model to effectively predict identity links, trained using the proposed latent node representation, selected knowledge graph features, and existing links.
\item We describe an algorithm for link discovery in the OSM datasets that uses the proposed supervised model and the latent representation to effectively identify missing links.
\end{itemize}

The results of the extensive experimental evaluation on three real-world OSM datasets for different geographic regions, along with the Wikidata and DBpedia knowledge graphs, confirm the effectiveness of the proposed \approach link discovery approach.
According to our evaluation results, \approach can reliably predict links. 

\approach achieves an F1 score of 92.05\% on Wikidata and of 94.17\% on DBpedia on average, which corresponds to a 21.82 percentage points increase in F1 score on Wikidata compared to the best performing baselines.

The remainder of the article is organised as follows. 
In Section \ref{sec:motivation}, we discuss the representation of geographic information in OSM and Wikidata and the existing interlinking between these sources to motivate our approach.
Then in Section \ref{sec:problem}, we formally introduce the link discovery problem addressed in this article. 
In Section \ref{sec:approach}, we present the proposed \approach approach.
Following that, we describe the evaluation setup in Section \ref{sec:setup} and provide and discuss our evaluation results in Section \ref{sec:evaluation}.
Then in Section \ref{sec:relatedWork}, we discuss related work.
Finally, in Section \ref{sec:conclusion}, we provide a conclusion.

\section{Motivation}
\label{sec:motivation}

\emph{Volunteered geographic information} is a special case of user-generated content that represents information about geographic entities \cite{Goodchild2007}. VGI is typically collected from non-expert users via interactive Web applications, with the OpenStreetMap project\footnote{\url{https://www.openstreetmap.org}} being one of the most prominent and successful examples. 
OSM is a rich source of spatial information available under an open license (Open Database License) and created collaboratively through an international community effort.
Today OSM data has become available at an unprecedentedly large scale. 
While in 2006 OSM captured only 14.7 million GPS points, this number has increased to 7.4 billion by 2019. Similarly the number of users who contribute to OSM has grown from 852 in 2006 to 5.6 million in 2019\footnote{\url{https://blackadder.dev.openstreetmap.org/OSMStats/}.}. 

\begin{table*}[!t]
    \footnotesize
    \centering
    \caption{Number of nodes, tags and distinct keys in the country-specific OSM snapshots (\texttt{OSM-[country]}) and their respective subsets linked to Wikidata (\texttt{Wikidata-OSM-[country]}).}

\resizebox{\textwidth}{!}{
\begin{tabular}{l@{\quad}rrr@{\hskip 1em}rrr@{\hskip 1em}rrr}
\toprule
        & \multicolumn{3}{c}{France} 
		& \multicolumn{3}{c}{Germany} 
		& \multicolumn{3}{c}{Italy}  \\
\cmidrule(l{0pt}r{6pt}){2-4} \cmidrule(l{0pt}r{6pt}){5-7} \cmidrule(l{0pt}r{0pt}){8-10}
 
    &  \texttt{OSM-FR} &  \texttt{Wikidata-OSM-FR} & Ratio
    &  \texttt{OSM-DE} &  \texttt{Wikidata-OSM-DE} & Ratio
    &  \texttt{OSM-IT} &  \texttt{Wikidata-OSM-IT} & Ratio\\
\midrule
 No. Nodes   & 390,586,064 & 21,629 & $0.01\%$
             & 289,725,624 & 24,312 & $<0.01\%$
             & 171,576,748 & 18,473 &  $0.01\%$  \\
No. Nodes with Name & 1,229,869 & 20,507  & $1.67\%$   
              & 1,681,481 & 23,979 & $1.43\%$
              & 557,189 & 18,420 & $3.31\%$ \\
 No. Tags   &  27,398,192 & 199,437 &  $ 0.73\% $ 
             & 37,485,549 &  212,727  &  $0.56\%$
             & 18,850,692 &  122,248 &  $0.65\%$ \\
 No. Distinct Keys   & 6,009 & 1,212  &  $20.17\%$
                    & 12,392 &  1,700 & $ 13.72\%$
                     & 4,349 & 892 & $20.51\%$ \\

\bottomrule
\end{tabular}
}
    \label{mot:tab:datasets}
\end{table*}

\begin{figure*}
    \centering
    \begin{subfigure}{0.32\textwidth}
        \includegraphics[width=\textwidth]{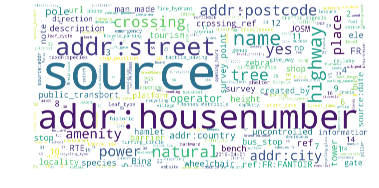}
        \caption{\texttt{OSM-FR}}
    \end{subfigure}
    \begin{subfigure}{0.32\textwidth}
        \includegraphics[width=\textwidth]{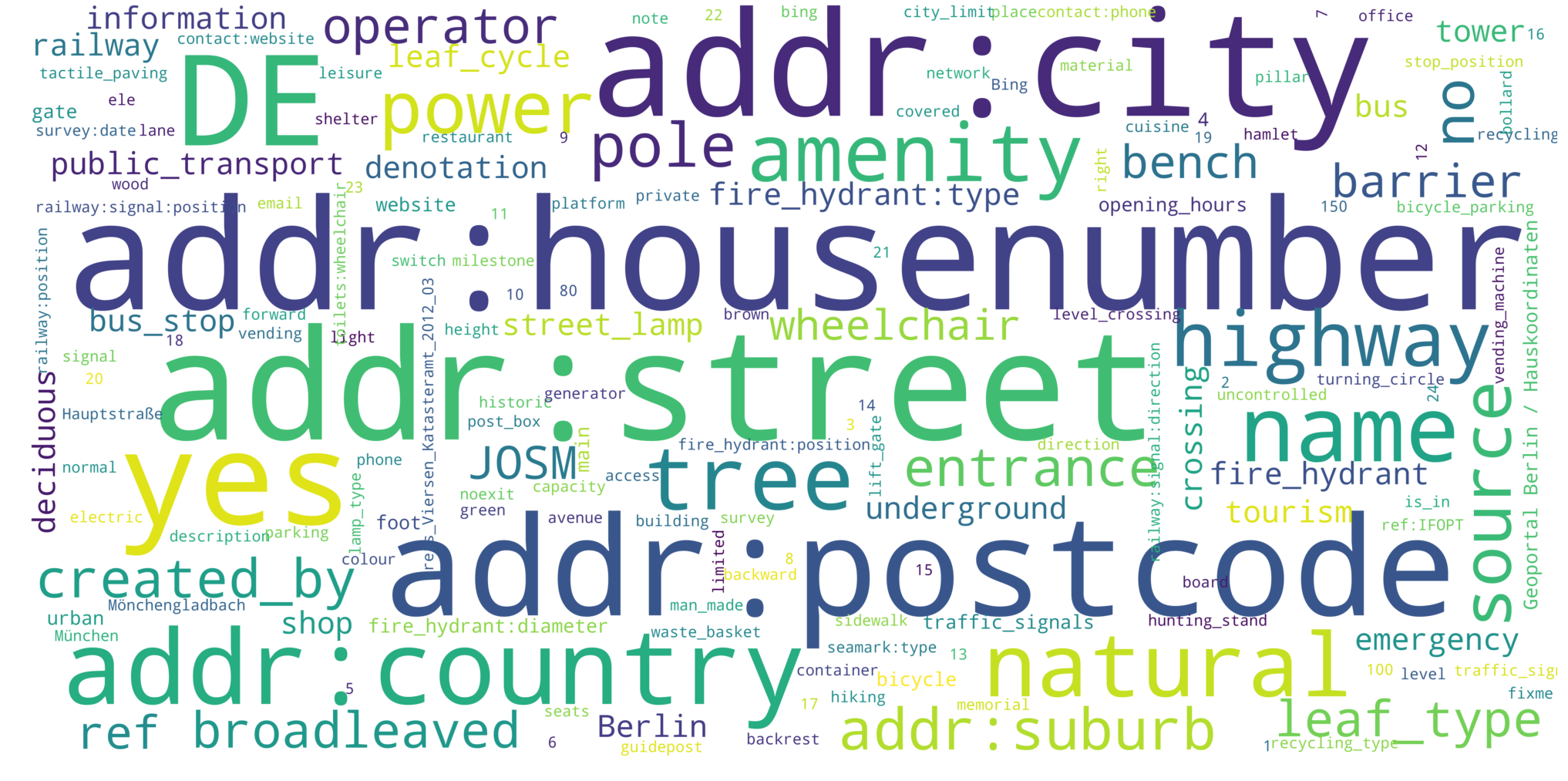}
        \caption{\texttt{OSM-DE}}
    \end{subfigure}
    \begin{subfigure}{0.32\textwidth}
        \includegraphics[width=\textwidth]{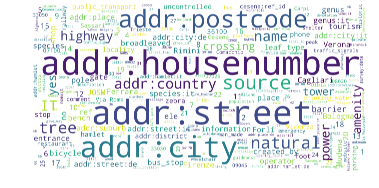}
        \caption{\texttt{OSM-IT}}
    \end{subfigure}
    \caption{Tag clouds generated from the 1000 most frequent tags in each respective OSM dataset.}
    \label{fig:tagcloud}
\end{figure*}

OSM includes information on \textit{nodes} (i.e., points representing geographic entities such as touristic sights or mountain peaks), as well as \textit{lines} (e.g. lists of points) and their topological \textit{relations}.
The description of nodes in OSM consists of few mandatory properties such as the node identifier and the location (provided as geographic coordinates) and an optional set of tags. 
\emph{Tags} provide information about nodes in the form of key-value pairs. 
For instance, the tag ``\texttt{place=city}'' is used to express that a node represents a city. 
OSM does not provide a fixed taxonomy of keys or range restrictions for the values but encourages its users to follow a set of best practices\footnote{\url{https://wiki.openstreetmap.org/wiki/Any\_tags\_you\_like}}. 
For example, the node labels are often available under the ``\texttt{name}'' key, whereas the labels in different languages can be specified using the ``\texttt{name:code=}'' convention\footnote{\url{https://wiki.openstreetmap.org/wiki/Multilingual\_names}}. 
The tags can also be used to specify identity links across datasets, e.g., to link OSM nodes to the equivalent entities in a knowledge graph.

For example, the link between the OSM node representing the city of Berlin  and its Wikidata counterpart is established via the tag ``\texttt{wikidata=Q64}'' assigned to the OSM node. Here, ``Q64''\footnote{\url{https://www.wikidata.org/wiki/Q64}} denotes the identifier of the corresponding Wikidata entity.
Recent studies indicate that the level of details provided for the individual OSM nodes is very heterogeneous \cite{Touya2015}. 
Contextual information, e.g., regarding the historical development of the city population, is typically not available in OSM. Furthermore, the individual keys and tags do not possess any machine-readable semantics, which further restricts their use in applications.

Country-specific OSM snapshots are publicly available\footnote{OSM snapshots can be found at \url{http://download.geofabrik.de}.}.
In the following, we refer to the country-specific snapshots as of September 2018 as the \texttt{OSM-[country]} dataset. E.g.m the snapshot for Germany is referred to as ``\texttt{OSM-DE}''. 
The linked sets \texttt{Wikidata-OSM-FR}, \texttt{Wikidata{\hyph}OSM{\hyph}DE}, and \texttt{Wikidata-OSM-IT} are the subsets of the \texttt{OSM{\hyph}[country]} datasets obtained by extracting all nodes that link to Wikidata entities from the respective OSM snapshot.  
Table \ref{mot:tab:datasets} provides an overview of the number of nodes, nodes with name,  tags, and distinct key contained in the \texttt{OSM{\hyph}[country]} datasets and the respective linked sets \texttt{Wikidata-OSM{\hyph}[country]}. As we can observe, only a small fraction of nodes, tags, and distinct keys from the overall datasets appear in the linked sets.
Furthermore, nearly all nodes contained in one of the linked sets exhibit a name tag.
In addition, in Figure \ref{fig:tagcloud}, we illustrate the most frequent keys 
of the \texttt{OSM-FR}, \texttt{OSM-DE}, and \texttt{OSM-IT} datasets
in a tag cloud visualisation.

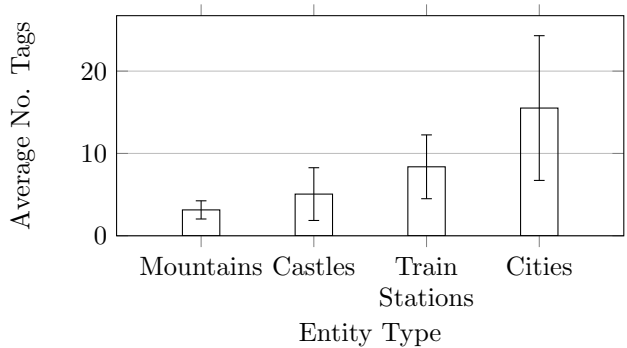
\begin{figure}[!t]
    \centering
    \begin{tikzpicture}

    \begin{axis}[
        width  = 0.45*\textwidth,
        height = 4.5cm,
        ybar=2*\pgflinewidth,
        bar width=14pt,
        ymajorgrids = true,
        ylabel = {Average No. Tags},
        xlabel = {Entity Type},
        symbolic x coords={Mountains, Castles, Train\\Stations, Cities},
        xtick = data,
        scaled y ticks = false,
        enlarge x limits=0.25,
        ymin=0,
        xticklabel style={align=center},
        xlabel style={yshift=-0.5cm},
        legend style={at={(0.02,0.9)},anchor=west, align=left}    ]
        \addplot[black, error bars/.cd,y dir=both,y explicit]
            coordinates {
            (Mountains, 3.143340) +- (Mountains, 1.104595)
            (Castles,5.063039) +- (Castles,3.203688) 
            (Train\\Stations, 8.377788) +- (Train\\Stations,3.874977)
            (Cities,15.513503) +- (Cities,8.789713)};

    \end{axis}
\end{tikzpicture}
    \caption{Average number of tags per entity type in \texttt{Wikidata-OSM-DE}. Error bars indicate the standard deviation.}
    \label{mot:fig:tagsPerType}
\end{figure}

Figure \ref{mot:fig:tagsPerType} depicts the mean and the standard deviation of the number of tags contained in the \texttt{OSM-DE} dataset for the four most common entity types in \texttt{Wikidata-OSM-DE}, such as cities, train stations, castles, and mountains. 
Note that, unlike a knowledge graph, OSM does not define the node type information explicitly. To generate the statistics presented in this section, we used the existing links between the OSM nodes and the Wikidata entities to manually identify the tags in OSM indicative for the particular entity types in Wikidata and collected the OSM nodes annotated with these tags.
We observe that the number of tags varies significantly with the entity type. Moreover the standard deviation is relatively high (between 35\% and 63\%)  for all entity types. While for some entity types (e.g., mountains) the variation in the absolute number of tags is rather small, other types (e.g., cities) exhibit more substantial variations, meaning that some of the cities possess more detailed annotations compared with the rest. 

\begin{figure}[!t]
    \centering
    \pgfplotsset{compat=1.11,
    /pgfplots/ybar legend/.style={
    /pgfplots/legend image code/.code={%
       \draw[##1,/tikz/.cd,yshift=-0.25em]
        (0cm,0cm) rectangle (8pt,1.2em);},
   },
}

\begin{tikzpicture}
    \setstretch{1}
  
    \begin{axis}[
        width  = 0.48*\textwidth,
        height = 5cm,
        ybar=5*\pgflinewidth,
        bar width=10pt,
        ymajorgrids = true,
        ylabel style={align=center},
        ylabel = {No. Nodes with Link\\ to Wikidata [\%]},
        xlabel = {Entity Type},
        symbolic x coords={Mountains, Castles, Train\\Stations, Cities},
        xtick = data,
        scaled y ticks = false,
        enlarge x limits=0.25,
        ymin=0,
        ymax=100,
        xticklabel style={align=center},
        legend style={at={(0.02,0.7)},anchor=west, align=left}
    ]
        \addplot[black,  postaction={pattern=north east lines}]
            coordinates {
            (Mountains, 11.506341)
            (Castles,52.799824)
            (Train\\Stations,67.362722)
            (Cities, 70.792079) };

        \addplot[black]
            coordinates {
            (Mountains, 5.373823)
            (Castles,36.695421)
            (Train\\Stations,31.065089)
            (Cities, 73.193010
) };
            
        \addplot[black,  postaction={pattern=grid}]
            coordinates {
            (Mountains, 6.941502)
            (Castles,13.802436)
            (Train\\Stations,78.262411)
            (Cities, 64.967105) };
    
        \legend{\texttt{OSM-FR}, \texttt{OSM-DE} ,\texttt{OSM-IT}}

    \end{axis}
\end{tikzpicture}
    \caption{Percentage of frequent OSM node types with links to Wikidata entities within the OSM datasets for Germany (\texttt{OSM-DE}), France (\texttt{OSM-FR}), and Italy (\texttt{OSM-IT}) as of September 2018.}
    \label{mot:fig:relLinks}
\end{figure}
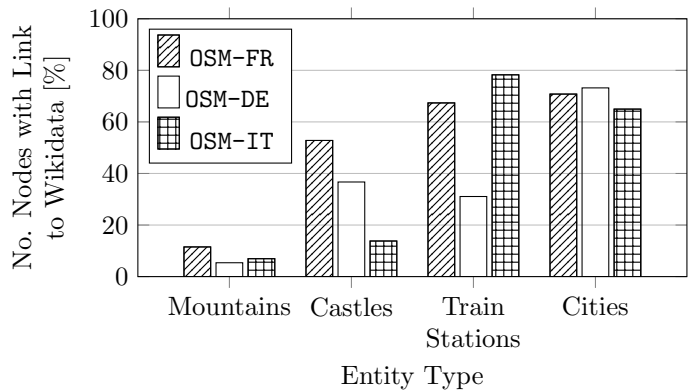

\begin{figure}[t]
    \centering
   \begin{subfigure}[t]{0.22\textwidth}
        \includegraphics[width=\textwidth]{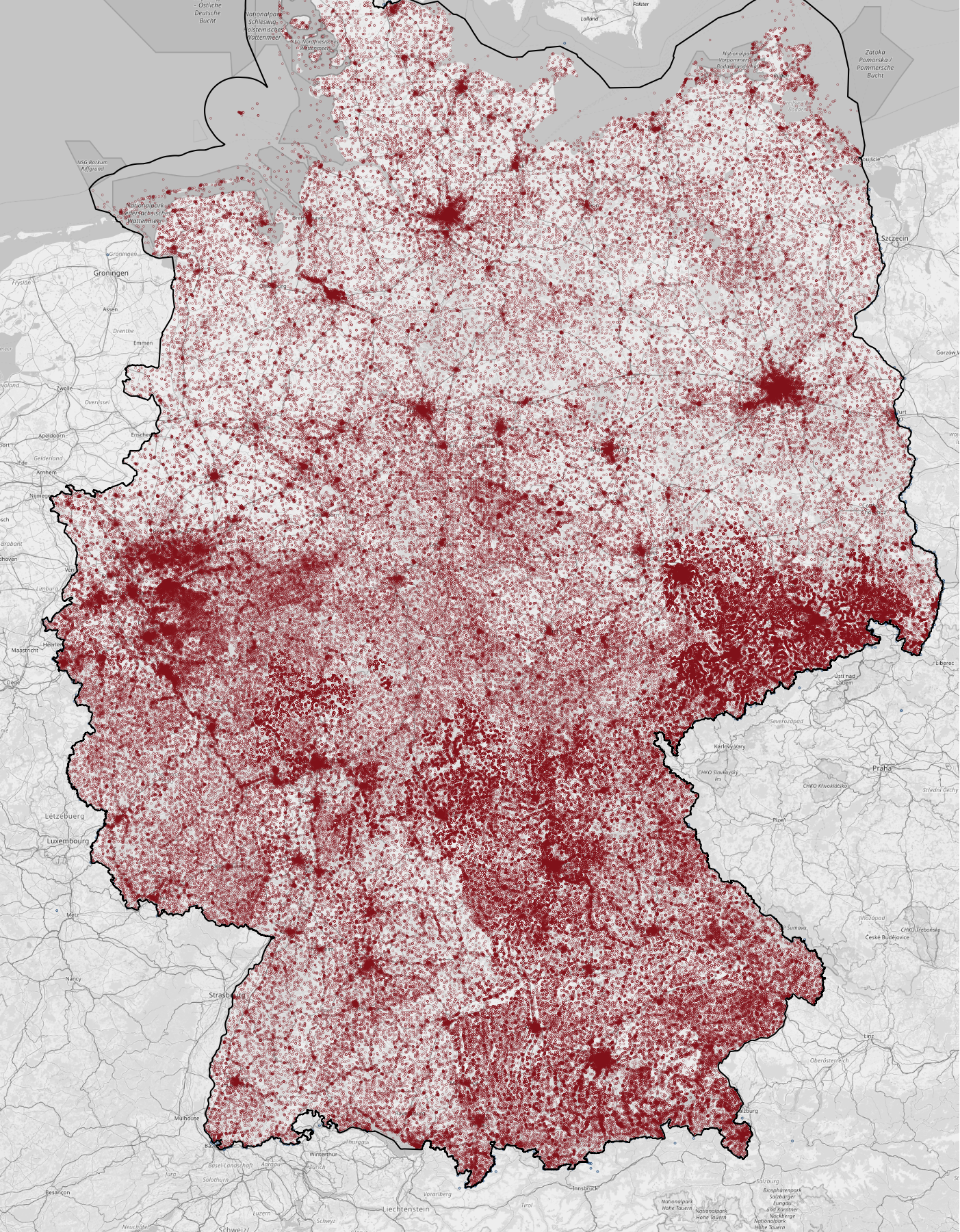}
                \caption{Wikidata}
    \end{subfigure}
    \begin{subfigure}[t]{0.22\textwidth}
        \includegraphics[width=\textwidth]{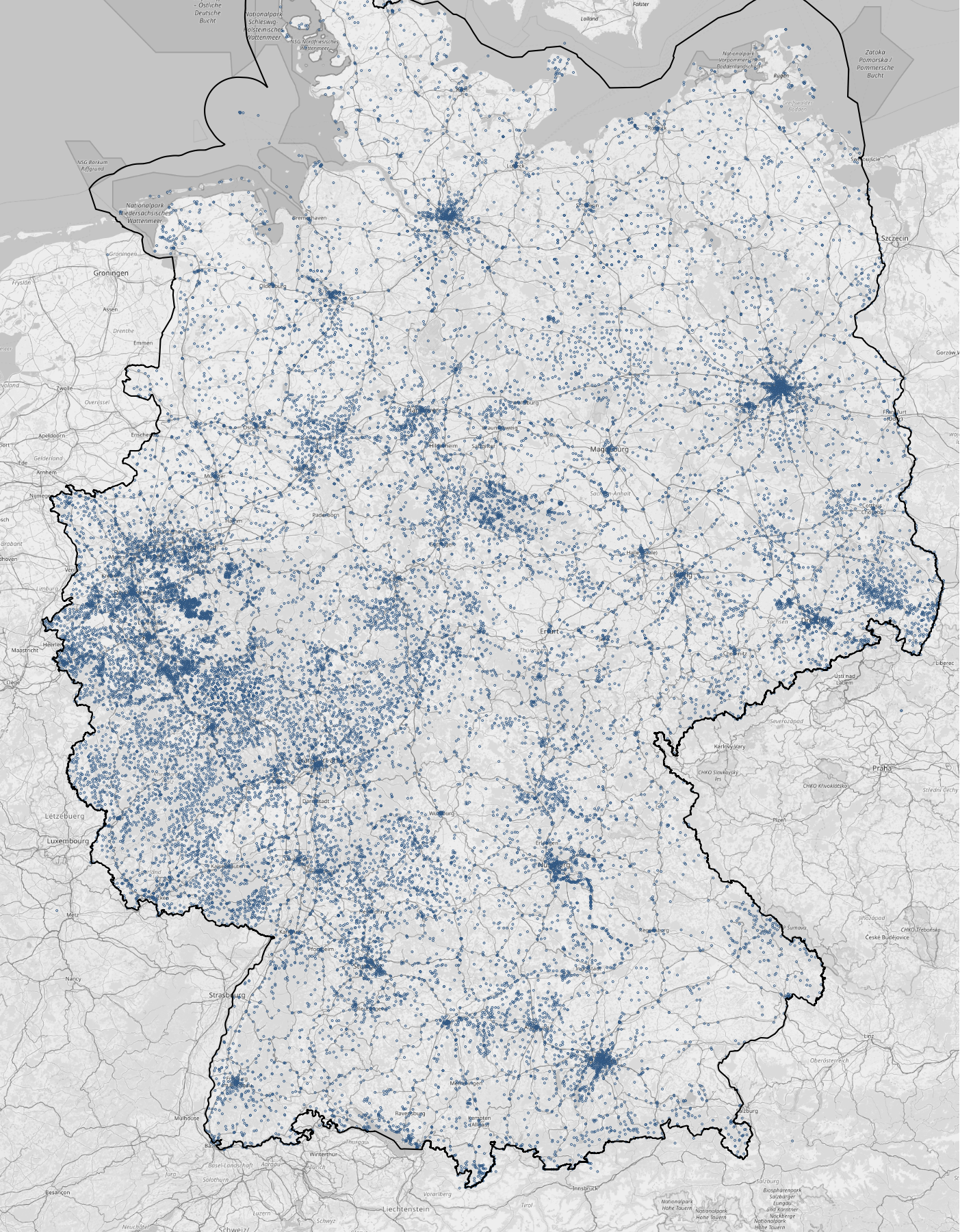}
        \caption{\texttt{Wikidata-OSM-DE}}
    \end{subfigure}
    \caption{Wikidata geo-entities located within Germany and Wikidata geo-entities linked by OSM. Map image: \textcopyright OpenStreetMap contributors, ODbL.}
    \label{mot:fig:wikidata_links}
\end{figure}

\emph{Knowledge graphs} such as Wikidata \cite{Vrandecic:2014}, DBpedia 
\cite{LehmannIJJKMHMK15}, 
and YAGO \cite{HOFFART2013} 
are a rich source of contextual information about geographic entities,
with Wikidata currently being the largest openly available knowledge graph linked to OSM.
In September 2018, Wikidata contained more than 6.4 million entities for which geographic coordinates are provided. 
Overall, the geographic information in OSM and contextual information regarding geographic entities in the existing knowledge graphs are highly complementary. 
As an immediate advantage of the existing effort to manually interlink OSM nodes and Wikidata entities, the names of the linked OSM nodes have become available in many languages \cite{GaneshBlog}. 

The links between the OSM nodes and geographic entities in Wikidata are typically manually annotated by volunteers and community efforts and are still only rarely provided. 
Figure \ref{mot:fig:relLinks} illustrates the percentage of the four most frequent geographic entity types (i.e., cities, train stations, mountains, and castles) that link to Wikidata from the OSM datasets for Germany, France, and Italy. 
Here, entity types are obtained from Wikidata using existing links between the OSM nodes and Wikidata entities.
As we can observe, the cities are linked most frequently, with a link coverage of approximately 70\% for all datasets. The link coverage of the other entity types is significantly lower, with mountains having the smallest coverage across these four categories with approximately 5\% in Germany. 
Figure \ref{mot:fig:wikidata_links} provides a visual comparison of the number of Wikidata entities located in Germany and the number of Wikidata entities to which links from OSM exist. 
While a significant fraction of links is still missing, existing links manually defined by volunteers reveal a high potential for being used as training data for supervised machine learning to increase link coverage automatically.%

In summary, volunteered geographic information is a continually evolving large-scale source of heterogeneous spatial data, whereas knowledge graphs provide complementary, contextual information for geographic entities. The links between VGI and knowledge graphs are mainly manually specified and are still only rarely present in the OSM datasets. The existing links represent a valuable source of training data for supervised machine learning methods to automatically increase the link coverage between OSM and knowledge graphs. This interlinking can provide a rich source of openly available semantic, spatial, and contextual information for geographic entities.

\section{Problem Statement}
\label{sec:problem}

In this work, we target the problem of identity link discovery between 
the nodes in a semi-structured geographic corpus such as OSM with equivalent
entities in a knowledge graph.

\begin{definition}
\textbf{Knowledge graph:}
Let $E$ be a set of entities, $R$ a set of labelled directed edges and $L$ a set of literals. 
A \emph{knowledge graph} $\mathcal{KG}=\langle E \cup L, R \rangle $ is a directed graph 
where entities in $E$ represent real-world entities and the edges in $R \subseteq (E \times E) \cup (E \times L)$ 
represent entity relations or entity properties. 
\label{def:knowledge_graph}
\end{definition}

In this work, we focus on interlinking entities in a knowledge graph 
that possess geographic coordinates, i.e., longitude and latitude. 
We refer to such entities as \textit{geo-entities}. Typical examples of geo-entities include cities, train stations, castles, and others.

\begin{definition}
\textbf{Geo-entity:}
A \emph{geo-entity} $e \in E$ is an entity for which a relation $r \in R$ exists that associates $e$ with geographic coordinates, i.e., a longitude $\textit{lon} \in L$ and a latitude $\textit{lat} \in L$.
\end{definition}

For instance, a geo-entity representing the city of Berlin may be represented as follows 
(the example illustrates an excerpt from the Wikidata representation of Berlin):

\[\left[
\begin{array}{lll}
    \textbf{Entity} &  \textbf{Property} & \textbf{Entity/Literal}\\
    Q64 & name & Berlin \\
    Q64 & instance ~of & Big ~City \\
    Q64 & coordinate ~location & 52\degree31'N, ~13\degree23'E \\
    Q64 & capital ~of & Germany\\
\end{array}\\[0.5em]
\right]\]

We denote the subset of nodes representing geo-entities in the knowledge graph $\mathcal{KG}$ as $E_{geo}\subseteq E$.  

\begin{definition}
\textbf{Geographic corpus:}
A \emph{geographic corpus} $\mathcal{C}$ is a set of nodes. 
A node $n \in \mathcal{C}$,  $n=\langle i, l, T \rangle$ is represented as a triple containing an identifier $i$, a location $l$, and a set of tags $T$. Each tag $t \in T$ is represented as a key-value pair with the key $k$ and a value $v$: $t=\langle k,v \rangle$. 
\end{definition}

For instance, the city of Berlin is represented as follows
(the example illustrates an excerpt from the OSM representation):

\[\left[
\begin{array}{ll}
    i & 240109189 \\
    l & 52.5170365, 13.3888599 \\
    \texttt{name=}& Berlin \\
    \texttt{place=} & city \\
    \texttt{capital=} & yes \\
\end{array}\\[0.5em]
\right]\]

Let $sameAs(n, e): \mathcal{C} ~\times ~E_{geo} \mapsto \{true, false\}$ be the predicate that holds iff $n\in \mathcal{C} $ and $e\in E_{geo}$ represent the same real-world entity. 
We assume that a node $n \in \mathcal{C}$ corresponds to at most one geo-entity in a knowledge graph $\mathcal{KG} $.
Then the problem of link discovery between a knowledge graph $\mathcal{KG} $ and a geographic corpus $\mathcal{C}$ is defined as follows.

\begin{definition}
\textbf{Link discovery:}
Given a node $n \in \mathcal{C}$ and the set of geo-entities $E_{geo}\subseteq E$ in the 
knowledge graph $\mathcal{KG}$, determine $e \in E_{geo}$ such that $sameAs(n, e)$ holds.
\end{definition}

In the example above, given the OSM node representing the city of Berlin, we aim to identify 
the entity representing this city in $E_{geo}$. 

\section{\approach Approach to Link Discovery}
\label{sec:approach}

The intuition of the proposed \approach approach is as follows:
\begin{itemize}
 \item[1.] Equivalent nodes and entities are located in geospatial proximity. Therefore, \approach adopts geospatial blocking to identify candidate entities in large-scale datasets efficiently.
\item[2.] OSM nodes are schema-agnostic and heterogeneous. 
Therefore \approach relies on an unsupervised model to infer latent, compact node representation that captures semantic similarity.
 \item[3.] Equivalent nodes and entities can indicate common representation patterns. Therefore, \approach adopts a supervised classification model for link prediction.
\end{itemize}

\begin{figure}
    \centering
    \includegraphics[width=0.48\textwidth]{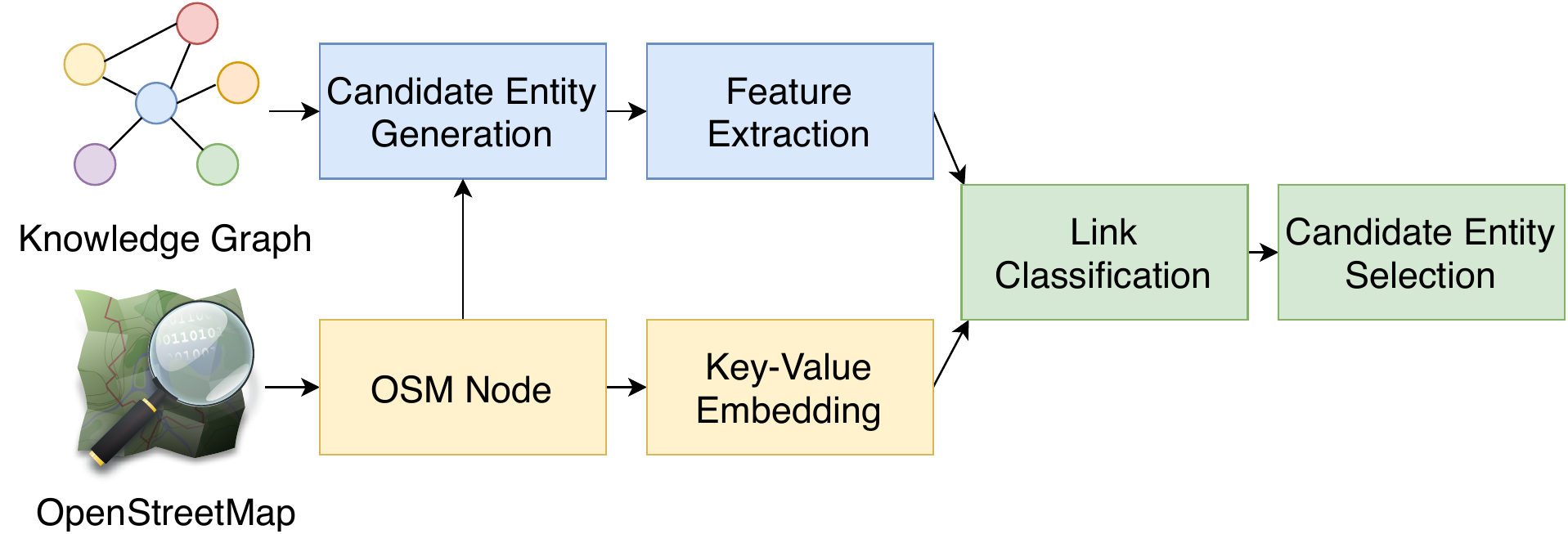}
    \caption{\approach Link discovery pipeline overview.}
    \label{fig:approach_overview}
\end{figure}

Figure \ref{fig:approach_overview} presents the \approach link discovery pipeline.
In the first blocking step, for each node $n \in \mathcal{C}$ in the geographic corpus $\mathcal{C}$, a set of candidates $E' \subseteq E_{geo}$ is generated from the set of geo-entities $E_{geo}$ contained in the knowledge graph.
In the next feature extraction step, representations of the node $n$ and the relevant entities $E'$ from the knowledge graph are extracted. 
A latent representation of the node $n \in \mathcal{C}$ is a \emph{key-value embedding} that is learned in an unsupervised fashion.
Representations of the knowledge graph entities in $E'$ are generated using selected knowledge graph features.
Furthermore, distance and similarity metrics for each candidate pair ($n \in \mathcal{C}$, $ e \in  E'$) are computed.
Following that, each candidate pair is processed by a supervised machine learning model during the link classification step. The model predicts if the pair represents the same real-world entity and provides a confidence score for the link prediction.
Finally, an identity link for the pair with the highest confidence among the positively classified candidate pairs for the node $n$ is generated. 
In the following, we discuss these steps in more detail.

\subsection{Candidate Entity Generation}
\label{sec:candidate_entity_generation}

Representations of a real-world geographic entity in different data sources may vary; this can be especially the case for the geographic coordinates in VGI, where the reference points represent typical points of community consensus rather than an objective metric \cite{SLHA11}. 
The blocking step is based on the intuition that geographic coordinates of the same real-world entity representation in different sources are likely to be in a short geographic distance.

Given a node $n \in \mathcal{C}$ contained in a geographic corpus and a knowledge graph $\mathcal{KG}=\langle E \cup L, R \rangle$, with a set of geo-entities ${E_{geo}\subseteq E}$, in the blocking step we compute a set of candidate geo-entities $E' \subseteq E_{geo}$ from $\mathcal{KG}$, i.e., the geo-entities potentially representing the same real-world entity as $n$.

The set of candidates $E'$ for a node $n$ consists of all geographic entities $e \in E_{geo}$ that are in a short geographic distance to $n$. In particular, we consider all entities within the distance specified by the blocking threshold $th_{block}$:
\[ E' = \{ e \in E_{geo}~ | ~\textit{distance}(n,e) \leq th_{block} \}, \]
where $distance(n,e)$ is a function that computes the geographic distance between the node $n$ and a geo-entity $e$. Here the geographic distance is measured as \emph{geodisc distance} \cite{387512}.

Note that $E'$ can be computed efficiently by employing spatial index structures such as R-trees \cite{Guttman84}. 
The value of the threshold $th_{block}$ can be  determined experimentally (see Section \ref{sec:blocking_threshold_evaluation}).


\subsection{Key-Value Embedding for Geographic Corpus}
\label{sec:approach:key-value-embedding}

In this work, we propose an unsupervised approach to infer novel latent representations of nodes in a geographic corpus. This representation aims at capturing the semantic similarity of the nodes by utilising typical co-occurrence patterns of OSM tags. 
Our approach is based on the intuition that semantic information, like for example entity types, can be inferred using statistical distributions \cite{PaulheimB14}. 
To realise this intuition in the context of a geographic corpus such as OSM, we propose a neural model inspired by the skip-gram model for word embeddings by Mikolov et al. \cite{NIPS2013_5021}. This model creates latent node representations that 
capture the semantic similarity of the nodes by learning typical co-occurrences of the OSM tags.

\begin{figure}
    \centering
    \includegraphics[width=0.48\textwidth]{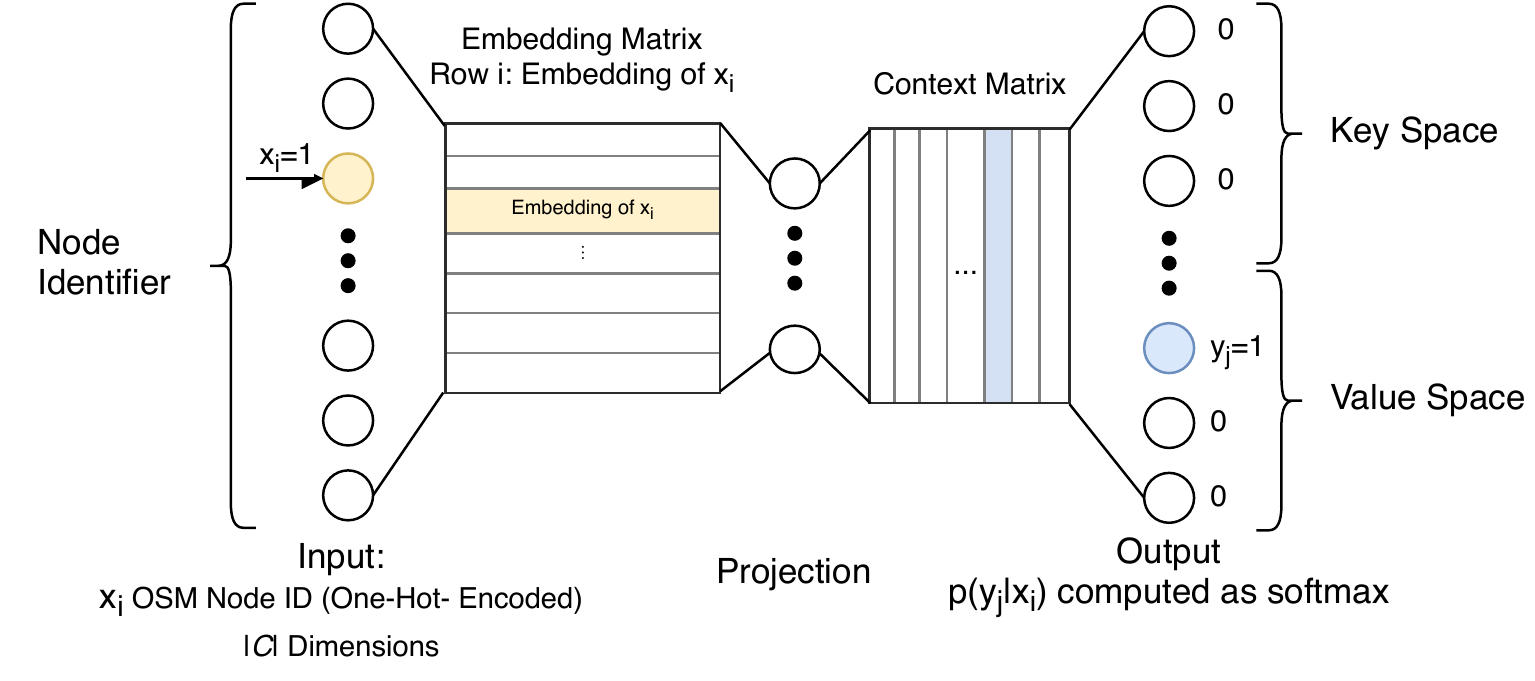}
    \caption{Architecture of the key-value embedding model. 
    The input layer 1-hot encodes the node identifiers.
    The embedding matrix transforms the input to the latent representation in the projection layer.
    The output layer maps the latent representation to the encoded keys and values by applying the softmax function.
  }
    \label{approach:fig:keyValEmbedding}
\end{figure}

In particular, we aim to obtain a latent representation of the node $n=\langle i,l,T \rangle, n\in \mathcal{C}$
that captures the semantic similarity of the nodes. 
To this extent, we propose a neural model that encodes the set of key-value pairs $T$ describing the node in an embedding representation. 
Figure \ref{approach:fig:keyValEmbedding} depicts the architecture of the adopted model that consists of an input, a projection, and an output layer. 
%
The \emph{input layer} encodes the identifier $n.i$ of each node $n=\langle i,l,T \rangle $. 
In particular, vector representations are obtained by applying one-hot-encoding\footnote{\url{https://www.kaggle.com/dansbecker/using-categorical-data-with-one-hot-encoding}} of the identifiers, i.e., each identifier $n.i$ corresponds to one dimension of the input layer. The corresponding entry of the vector representation is set to 1, while other entries are set to 0.  
The \emph{projection layer} computes the latent representation of the nodes. 
The number of neurons in this layer corresponds to the number of dimensions in the projection, i.e., the embedding size.
The \emph{output layer} maps the latent representation to the encoded keys and values using softmax \cite{Goodfellow-et-al-2016}. The key-value pairs $\langle k,v \rangle \in n.T$ for each node $n$ are encoded by applying one-hot-encoding to both keys and values separately. As the set of values might be highly diverse, we only consider the top-k most frequent values to be represented as an individual dimension. 
The non-frequent values are unlikely to be indicative for semantic similarity, whereas the information of the presence of a rare value can be discriminative. Thus, all non-frequent values are mapped to a single dimension.

The embedding aims to generate a similar representation for the nodes with similar properties, independent of their location. Therefore, we do not include location information, such as geographic coordinates, in the embedding. Note that the value of name tags are typically not part of the embedding, as names typically have rare values.

The objective of the proposed model is to maximise the following log probability:
\[ \sum_{n \in \mathcal{C}} \sum_{\langle k,v \rangle \in n.T} \log p(k|n.i)+\log p(v|n.i).\]

Here, the term $\log p(k|n.i) + \log p(v|n.i)$ expresses the node's log probability with the identifier $n.i$ to be annotated with the key-value pair $\langle k,v \rangle$, i.e. $\langle k,v\rangle \in n.T$. The probabilities are calculated using softmax. 
The training of the network aims at minimising the key-value based loss function. 
This way, nodes that exhibit similar keys or values are assigned similar representations in the projection layer. 
Thus, we use the activation of the projection layer as a latent representation of each respective OSM node. This representation captures the latent semantics of the keys and values of the node.
We refer to this feature as \textit{KV-embedding}. We learn the \textit{KV-embedding} for each OSM node.
The training is conducted without any supervision. 
The resulting node representation is task-independent.

\subsection{Feature Extraction from KG}
\label{sec:feature extraction}

This step aims at extracting features for the entities $e \in E'$, where $E'$ denotes the set of candidate geo-entities in the knowledge graph for the target node $n \in \mathcal{C}$. 
We adopt the following features: 

\emph{Entity Type}: Entities and nodes that belong to the same category, for instance ``city'' or ``train station'', are more likely to refer to the same real-world entity than the candidates of different types. 
In the knowledge graph, we make use of the \textit{rdf:type}\footnote{rdf: \url{http://www.w3.org/1999/02/22-rdf-syntax-ns}} property as well as knowledge graph specific properties (e.g. \textit{wikidata:instanceOf}) to determine the type of $e$. 
To encode the type, we create a vector of binary values in which each dimension corresponds to an entity type. For each type of $e$, the corresponding dimension is set to ``1'' while all other dimensions are set to ``0''.
Concerning the target node $n$, the node type is not expected to be explicitly provided in a geographic corpus. 
Nevertheless, we expect that the \textit{KV-embedding} of the geographic corpora implicitly encodes type information, based on the intuition that types can be inferred using statistical property distributions \cite{PaulheimB14}. 

\emph{Popularity}: A similar level of entity popularity in the respective sources can provide an indication for matching. 
Popular entities are likely to be described with a higher number of relations 
and properties than less popular entities. To represent entity popularity, 
we employ the number of edges starting from $e$ in $\mathcal{KG}$ as a feature. 
More formally: $\textit{popularity}(e)=|\{(e,x) \in R\ ~| ~x \in E \cup L\}|$.
We expect that the \textit{KV-embedding} implicitly encodes the node popularity information in the geographic corpora as popular nodes have a higher number of tags.

\subsection{Similarity and Distance Metrics}
\label{sec:distance-and-similarity}

This step aims at extracting features that directly reflect the similarity
between an OSM node $n \in \mathcal{C}$ and a candidate geo-entity $e \in E'$.
To this extent, we utilise name similarity and geographical distance.

\emph{Name Similarity}:
Intuitively, a geo-entity and an OSM node sharing the same name are likely to represent the same real-world object. 
Therefore, we encode the similarity between the value of the \textit{name} tag of an OSM node $n \in \mathcal{C}$ and the \textit{rdfs:label}\footnote{rdfs: \url{http://www.w3.org/2000/01/rdf-schema}} of a geo-entitiy $e \in E'$ as a feature. We compute the similarity using the Jaro-Winkler distance \cite{winkler99}, also adopted by \cite{SLHA11}. 
Jaro-Winkler distance assigns a value between [0,1], where 0 corresponds to no difference and 1 to the maximum dissimilarity.
If a \textit{name} tag or a \textit{rdfs:label} is not available for a particular pair $(n, e)$, the value of this feature is set to 1.

\emph{Geo Distance}: Based on the intuition that nodes and candidate entities that exhibit smaller geographic distance are more likely to refer to the same real-world entity, we employ geographic distance as a feature. 
To this extent, we utilise the logistic distance function proposed in \cite{SLHA11}:
\[
\textit{geo-distance}(n,e) = 1/(1+exp(-12d'(n,e)+6)),
\]
with $d' = 1 - d(n,e)/th_{block}$, where $d$ denotes the so-called \emph{geodisc distance} \cite{387512} between $n$ and $e$ and takes the spheroid form of the earth into account. 
$th_{block}$ denotes the threshold that defines the maximum geographic distance at which the candidates are considered to be similar.
To facilitate efficient computation, the $th_{block}$ threshold is also utilised in the blocking step, described in Section \ref{sec:candidate_entity_generation}.
The intuition behind the logistic distance function is to allow for smaller differences of the geographic positions and to punish more significant differences.
The Geo Distance feature directly encodes the geospatial similarity between the node $n$ and the candidate geo-entity $e$.

\subsection{Link Classification}
\label{sec:link_classification}

We train a supervised machine learning model to predict whether the target node $n \in \mathcal{C}$ and a candidate geo-entity represent the same real-world entity. 
Each target node $n$ and the set of candidates $E'$ for this node 
are transformed into the feature space. 
Each node-candidate pair is interpreted as an instance for a supervised machine learning model by concatenating the respective feature vectors. 
For training, each pair is then labelled as correct or incorrect, where labels are obtained from the existing links to the knowledge graph within the OSM corpus $\mathcal{C}$.
Note that the number of pairs labelled as incorrect (i.e., negative examples) is typically higher than the number of correct pairs. 
To allow an efficient training of classification models, we limit the number of incorrect candidates for each node $n$ to 10 candidates via random sampling.  
To address the imbalance of classes within the training data, we employ oversampling to level out the number of instances per class. In particular, we employ the state-of-the-art SMOTE algorithm \cite{DBLP:journals/jair/ChawlaBHK02}. 
The data is then normalised by removing the mean and scaling to unit variance. 
We use the normalised data as input to the classification model. 
We consider the following models: \textsc{Random Forest}, \textsc{Decision Tree}, \textsc{Na\"ive Bayes}, and \textsc{Logistic Regression}. 
We discuss the model performance in Section \ref{sec:classifier_evaluation}.
We optimise the hyperparameters using random search \cite{Bergstra:2012:RSH:2188385.2188395}. 

Finally, the candidate entity selection is based on the assumption that the knowledge graph contains at most one geo-entity equivalent to the target node.
If at least one node within $E'$ is classified as correct (with a confidence $>50\%$), a link between node $n$ and $e_{max} \in E'$ is created, where $e_{max}$ denotes the entity with the highest confidence score of the model. 
If all entities are labelled as incorrect, no link for the node $n$ is created.

\subsection{Algorithm for Link Discovery}
Finally, Algorithm \ref{alg:linkDiscov} details the process of link discovery. The algorithm integrates the above described steps, namely \emph{candidate entity generation} (line 1), \emph{feature extraction} (lines 2-7), \emph{link classification} (lines 9-12) and \emph{candidate entity selection} (lines 12-17). Table \ref{tab:functions} presents a description of the functions used in the algorithm.

\begin{algorithm}
    \small
\begin{flushleft}
	\begin{tabular}{lll}
	\texttt{Input}: & Node $n \in \mathcal{C}$ &\\
	                & Knowledge graph $\mathcal{KG}$ &\\
 	\texttt{Output}: & Entity $e_{link} \in \mathcal{KG}$ that should be linked to $n$ &\\
 	                & or \texttt{null} if no matching entity was found  & 
	\end{tabular}

\end{flushleft}
\begin{algorithmic}[1]
	\STATE $E' \leftarrow \texttt{generateCandidates}(n, \mathcal{KG})$
    
    \STATE features $\leftarrow$ []
     \STATE features[$n$] $\leftarrow$ \texttt{KV-embedding}($n$)
     \FORALL{$e \in E'$} 
        \STATE{\text{features}[$e$] $\leftarrow$ \texttt{KG-features($e$, $\mathcal{KG}$)}}
        \STATE{\text{features}[$e$] $\leftarrow$ \text{features}[$e$] $\cup$ \texttt{similarity-features($e$, $n$)}}

    \ENDFOR
    
    \STATE confidences $\leftarrow$ []
    \FORALL{$e \in E'$} 
        \STATE{confidences[$e$] $\leftarrow$ \texttt{link-classification}(features[$n$], features[$e$])}
    \ENDFOR
    \STATE $e_{link}$ $\leftarrow \argmax_{e \in E'}$(confidences[$e$])

    \IF{\texttt{classifieddAsCorrect}($e_{link})$}
        \RETURN $e_{link}$
    \ELSE
        \RETURN \texttt{null}
    \ENDIF
    
\end{algorithmic}
\caption{Link Discovery}
\label{alg:linkDiscov}
\end{algorithm}

\begin{table}
\caption{Description of functions used in Algorithm \ref{alg:linkDiscov}.}
\footnotesize
\begin{tabularx}{0.48\textwidth}{lXc}
    
    \toprule
    \textbf{Function Name} & \textbf{Returned Result} & \textbf{Section} \\
    \midrule
     \texttt{generateCandidates} & Candidate entities from $\mathcal{KG}$ nearby $n$   & \ref{sec:candidate_entity_generation} \\
     \texttt{KV-embedding} & Latent representation of $n$ & \ref{sec:approach:key-value-embedding}\\
     \texttt{KG-features} & Feature representation for $e$  & \ref{sec:feature extraction} \\
     \texttt{similarity-features} & Similarity features between $e$ and $n$ & \ref{sec:distance-and-similarity} \\
     \texttt{link-classification} & Confidence score for ($n, e$) & \ref{sec:link_classification}\\
     \texttt{classifiedAsCorrect} & True iff a link between ~($n, e$) is classified to be correct  & \ref{sec:link_classification}\\
     \bottomrule
\end{tabularx}
\label{tab:functions}
\end{table}

\subsection{Implementation}
\label{sec:implementation}
In this section, we provide implementation details of the \approach components.
We implemented our overall experimental framework and the proposed algorithm in Java 8. 
We stored the evaluation results in a PostgreSQL\footnote{\url{https://www.postgresql.org/}} database (version 9.6).
In a pre-processing step, we extracted relevant data from OpenStreetMap using Python (version 3.6) and the osmium\footnote{\url{https://osmcode.org/libosmium/}} library (version 2.14).
We extracted relevant knowledge graph entities from Wikidata with geographic coordinates using pyspark\footnote{\url{https://spark.apache.org/docs/latest/api/python/pyspark.html}} (version 2.2).
The geographic data was stored in a PostgreSQL database (version 9.6) and indexed using the PostGIS\footnote{\url{https://postgis.net/}} extension (version 2.3).
The feature extraction is implemented in Java 8 within our experimental framework.
We implemented the extraction of the KV-embedding in Python 3.6, using Tensorflow\footnote{\url{https://www.tensorflow.org/}} version 1.14.1.
The machine learning algorithms were implemented in Python 3.7 using the scikit-learn\footnote{\url{https://scikit-learn.org/stable/}} (version 0.21) and the imbalanced-learn\footnote{\url{https://imbalanced-learn.readthedocs.io/en/stable/api.html}} (version 0.5) libraries. 
To facilitate the reproducibility, we make our code available under the open MIT license in a GitHub repository\footnote{\url{https://github.com/NicolasTe/osm2kg}}.

\section{Evaluation Setup}
\label{sec:setup}

In this section, we describe the datasets, metrics, baselines and \approach configurations utilised in the evaluation. 

\subsection{Datasets and Metrics}
\label{sec:datasets}

We conduct our evaluation on three large-scale OSM datasets for France, 
Germany, and Italy as well as the Wikidata and DBpedia knowledge graphs. 

\emph{Knowledge Graphs}: In our experiments, we consider Wikidata snapshot from September 2018, as well as DBpedia in its German, French and Italian editions, snapshots from August 2019, as the target knowledge graphs.
\textit{Wikidata} \cite{Vrandecic:2014} is a publicly available collaborative knowledge graph. Wikidata is the central repository for structured information of the Wikimedia Foundation and the currently largest openly available knowledge graph.
\textit{DBpedia} \cite{LehmannIJJKMHMK15} is a knowledge graph that extracts structured data from the information of various Wikimedia projects, e.g., the Wikipedia\footnote{\url{https://www.wikipedia.org}} encyclopedia. DBpedia is provided in language-specific editions.
We refer to each language-specific edition of DBpedia as \texttt{DBpedia-[language]}.
Table \ref{tab:setup:kg_stats} presents the number of available geographic entities as well as the number of distinct types and the average number of edges per geo-entity in each knowledge graph.
Note that we consider geo-entities in the knowledge graphs with valid geographic coordinates, i.e., coordinates that can be located on the globe.

\begin{table}
    \caption{The number of geographic entities, distinct types and average statements per geo-entity in the considered knowledge graphs.}
    \label{tab:setup:kg_stats}
    \footnotesize
    \begin{tabular}{lrrr}
        \toprule
        \makecell{\textbf{Knowledge}\\\textbf{Graph}} & \makecell{\textbf{No.}\\\textbf{Geo-Entities}} &  \makecell{\textbf{No. Distinct}\\\textbf{Types}} & \makecell{\textbf{Average No.}\\\textbf{Edges/Entity}} \\
        \midrule
        \texttt{Wikidata} & 6,465,081 & 13,849  & 24.69  \\
        \texttt{DBpedia-FR} & 317,500 & 185 & 18.33 \\
        \texttt{DBpedia-DE} & 483,394 & 129 & 31.60 \\
        \texttt{DBpedia-IT} & 111,544 & 11 & 31.13 \\
         \bottomrule 
    \end{tabular}
\end{table}

\emph{OpenStreetMap:} 
We consider OSM datasets extracted from the three largest country-specific OSM snapshots as of September 2018. 
In particular, we consider the snapshots of Germany, France, and Italy.
We denote the country-specific snapshots as \texttt{OSM-[country]}.
Furthermore, we extract all nodes that exhibit a link to a geo-entity contained in Wikidata or DBpedia.
For DBpedia, we consider links to the DBpedia version of the language that corresponds to the country of the individual OSM snapshot, since the existing links in the country-specific snapshots target the respective language-specific edition of DBpedia in all cases for the considered datasets.
We denote the considered link datasets as \texttt{[KG]-OSM-[language]}. 
For instance, \texttt{DBpedia-OSM-FR} denotes the dataset that interlinks the OSM snapshot of France with the French DBpedia.

\begin{table}
    \caption{The number of existing links between OpenStreetMap, Wikidata and DBpedia. \texttt{OSM-[country]} denote the country-specific snapshots of OSM as of September 2018. The existing links serve as ground truth for the experimental evaluation.}
    \label{tab:setup:kg_links}
    \footnotesize
    \centering
    \begin{tabular}{lrrr}
        \toprule
        \textbf{Knowledge Graph} & \texttt{OSM-FR} &  \texttt{OSM-DE} &  \texttt{OSM-IT} \\
        \midrule
        \texttt{Wikidata} & 21,629  & 24,312 & 18,473 \\
        \texttt{DBpedia-FR} & 12,122 & - & - \\
        \texttt{DBpedia-DE} & - & 16,881 & - \\
        \texttt{DBpedia-IT} & - & - & 2,353 \\
         \bottomrule 
    \end{tabular}
\end{table}

Table \ref{tab:setup:kg_links} provides an overview of the number of existing links between OSM and the knowledge graphs.
The existing links between the OSM datasets and knowledge graphs in these link datasets serve as ground truth for the experimental evaluation of all link discovery approaches considered in this work.

To assess the performance of link discovery approaches, we compute the following metrics:

	\textbf{Precision}: The fraction of the correctly linked OSM nodes among
	all nodes assigned a link by the considered approach.
	
	\textbf{Recall}: The fraction of the OSM nodes correctly linked by the approach among all
	nodes for which links exist in the ground truth.
	
	\textbf{F1 score}: The harmonic mean of recall and precision.
In this work, we consider the F1 score to be the most relevant metric since it reflects both recall and precision. 

We apply the 10-fold cross-validation. We obtain the folds by random sampling the links from the respective link datasets. For each fold, we train the classification model on the respective training set. We report the macro average over the folds of each metric.

\subsection{Baselines}
\label{sec:baselines}

We evaluate the link discovery performance of \approach\linebreak
against the following unsupervised and supervised baselines:

\textbf{\textsc{BM25}}: This naive baseline leverages the standard BM25 text retrieval model \cite{manning2008} to predict links. We created an inverted index on English labels of all geo-entities (i.e., for all $e \in E_{geo}$)  in a pre-processing step to apply this model. Given the target node $n$, we query the index using the value of the name tag of $n$ to retrieve geo-entities with similar labels.
We query the index using either the English name tag of the node $n$ (if available) or the name tag without the language qualifier. We create the link between $n$ and the entity with the highest similarity score returned by the index. If the name tag is not available, we do not create any link.

\textbf{\textsc{SPOTLIGHT}}:
This baseline employs the \emph{DBpedia Spotlight} \cite{Daiber:2013:IEA:2506182.2506198} model to determine the links. 
\emph{DBpedia Spotlight} is a state-of-the-art model to perform entity linking, i.e., to link named entities mentioned 
in the text to the DBpedia knowledge graph.
Given an OSM node $n$, we use the name tag of this node in the language native to the specific OSM dataset as an input to the DBpedia Spotlight model in the same language edition. 
The model returns a set of DBpedia entities out of which we choose the entity with the highest confidence score. 
\textbf{}To increase precision, we restrict the DBpedia Spotlight baseline to return only entities of type \textit{dbo:Place}\footnote{dbo: DBpedia Ontology}.
DBpedia entities are resolved to the equivalent Wikidata entities using existing \textit{wikidata:about} links.

\textbf{\textsc{GEO-DIST}}:
This baseline predicts the links solely based on the geographic distance, measured as geodisc distance. For a target OSM node $n$, the link is created between $n$ and $e_{min} \in E_{geo}$, where 
\[
e_{min} = \textit{argmin}_{e \in E_{geo}}(\textit{distance}(n,e)). 
\]
Here, $\textit{distance}(n,e)$ is a function that computes the geodisc distance between the OSM node $n$ and the geo-entity $e$.

\textbf{\textsc{LGD}}: This baseline implements a state-of-the-art approach of interlinking OSM with a knowledge graph proposed in the \emph{LinkedGeoData} project \cite{SLHA11}.
The \textsc{LGD} baseline utilises a combination of name similarity computed using the \emph{Jaro–Winkler} string distance
and geographic distance. It aims at computing links with high precision. 
For each OSM node $n$ a link between $n$ and $e \in E_{geo}$ is generated if the condition $\frac{2}{3}s(n,e)+\frac{1}{3}g(n,e, th_{block}) > th_{str}$ is fulfilled, where $th_{str} = 0.95$
Here, $s(n,e)$ denotes the Jaro-Winkler distance between the value of the name tag of $n$ and the label of $e$. 
If the name tag is not available, an empty string is used to compute the distance.
$g(n,e, th_{block})$ is a logistic geographic distance function specified in \cite{SLHA11}. 
The parameter $th_{block}$ denotes the maximum distance between a geo-entity and the node $n$. 
In our experiments, we use $th_{block}= 20000$ meter to allow for high recall.

\textbf{\textsc{LGD-SUPER}}: We introduce supervision into the \textsc{LGD} baseline by performing exhaustive grid search for $th_{block} \in \{1000, 1500, 2500, 5000, 10000, 20000\}$ meter and $th_{str}  \in \{ 0.05 \cdot i ~| ~i \in \mathbb{N}, 1 \leq i \leq 20  \}$. We evaluate each combination on the respective training set and pick the combination that results in the highest F1 score.

\textbf{\textsc{YAGO2GEO}}: This method was proposed in \cite{10.1007/978-3-030-30796-7_12}
to enrich the YAGO2 knowledge graph with geospatial information from external sources, including OpenStreetMap.
Similar to \textsc{LGD}, this baseline relies on a combination of the Jaro-Winkler and geographic distance. 
In particular, a link between an OSM node $n$ and $e \in E_{geo}$ is established if $s(n,e) < th_{str}$ and $distance(n,e) < th_{block}$ with $th_{str}=0.82$, $th_{block} = 20000$ meter. $s(n,e)$ denotes the Jaro-Winkler distance between the value of the name tag of $n$ and the label of $e$, and $distance(n,e)$ denotes the geographic distance between $e$ and $n$.

\textbf{\textsc{YAGO2GEO-SUPER}}: We introduce supervision into the \textsc{YAGO2GEO} baseline by performing exhaustive grid search for $th_{block} \in \{1000, 1500, 2500, 5000, 10000, 20000\}$ meter and $th_{str}  \in \{ 0.05 \cdot i ~| ~i \in \mathbb{N}, 1 \leq i \leq 20  \}$. We evaluate each combination on the respective training set and pick the combination that results in the highest F1 score.

\textbf{\textsc{LIMES/Wombat}}: The Wombat algorithm, integrated within the LIMES framework \cite{10.1007/978-3-319-58068-5_7}, is a state-of-the-art approach for link discovery in knowledge graphs. 
The algorithm learns rules, so-called link specifications, that rate the similarity of two entities.
The rules conduct pairwise comparisons of properties, which are refined and combined within the learning process. 
As LIMES requires the data in the RDF format, we transformed the OSM nodes into RDF triples, in which the OSM id represents the subject, the key represents the predicate, and the value represents the object.
We further added \textit{geo:lat}\footnote{geo: \url{http://www.w3.org/2003/01/geo/wgs84\_pos}} and \textit{geo:long} properties representing geographic coordinates of the OSM nodes.
LIMES requires all entities to contain all considered properties. 
Therefore we limit the properties to the geographic coordinates \textit{geo:lat}, \textit{geo:lon} as well as the name tag in OSM and the \textit{rdfs:label}\footnote{rdfs: \url{http://www.w3.org/2000/01/rdf-schema}} in the knowledge graph.
We use the default similarity metrics of LIMES, namely Jaccard, trigram, 4-grams, and cosine similarity and accept all links with a similarity score higher or equal to 0.7. 
Note that LIMES does not distinguish between data types when using machine learning algorithms.
Therefore, it is not possible to simultaneously use string similarity and spatial similarity metrics (e.g. Euclidean distance).

\subsection{\approach Configurations}
\label{sec:training_of_key_value_embeddings}

We evaluate our proposed \approach approach in the following configuration:
\textsc{Random Forest} as classification model (according to the results presented later in Section \ref{sec:classifier_evaluation}, \textsc{Random Forest} and \textsc{Decision Tree} perform similarly on our datasets), dataset-specific embedding size of 3-5 dimensions (Section \ref{sec:embedding_size_evaluation}),
and a blocking threshold of 20 km for \texttt{DBpedia-OSM-IT} and
2.5 km for all other datasets (Section \ref{sec:blocking_threshold_evaluation}). 

Furthermore, we evaluate our proposed approach in the following variants:

\noindent \textbf{\textsc{OSM2KG}}:
In this variant, we run \approach as described in Section \ref{sec:approach} using the features KV-embedding, Name Similarity, Geo Distance, Entity Type, and Popularity.
To obtain latent representations of the OSM nodes, we train unsupervised embedding models as described in Section \ref{sec:approach:key-value-embedding} on each of the \texttt{OSM-FR}, \texttt{OSM-IT}, \texttt{OSM-DE} datasets. 
During training, we consider the top-k most frequent values with k=1000 to be represented in the value space and compute 1000 epochs using a learning rate of $\alpha=1.0$. 
We make the key-value embeddings of OpenStreetMap nodes created in our experiments publicly available\footnote{\url{http://l3s.de/~tempelmeier/osm2kg/key-value-embeddings.zip}}.
These key-value embeddings provide a task-in\-de\-pen\-dent compact representation of OSM nodes. 

\noindent \textbf{\textsc{OSM2KG-TFIDF}}:
To better understand the impact of the proposed embedding method on the link discovery performance, in this variant, we exchange the proposed KV-embedding with a simple TF-IDF representation of the keys and values (i.e., term frequency and inverse document frequency). 
To this extent, we computed the TF-IDF values of the top 1000 most frequent keys and values for each OSM dataset. 
In this representation, each of the keys and values is described by a single dimension, resulting in a 1000-dimension vector.
All other features, such as Name Similarity, Geo Distance, Entity Type, and Popularity remain the same.

\section{Evaluation}
\label{sec:evaluation}

\begin{table*}
    \centering

    \caption{Macro averages for precision, recall and F1 score [\%], best scores are bold. Statistically significant (according to paired t-tests with $p < 0.05$) F1 score results of \textsc{OSM2KG} compared to all baselines and \textsc{OSM2KG-tfidf} are marked with *.}
    \begin{subtable}[c]{\textwidth}
    \footnotesize
   \caption{Link prediction performance on the Wikidata datasets}
    \label{ev:tab:perfomanceWikidata}
    \begin{tabular}{l@{\quad}rrc@{\hskip 1em}rrc@{\hskip 1em}rrc@{\hskip 1em}rrc}
\toprule
\multirow{2}{*}{\raisebox{-\heavyrulewidth}{\textbf{Approach}}} &               \multicolumn{3}{c}{\texttt{Wikidata-OSM-FR}} 
		& \multicolumn{3}{c}{\texttt{Wikidata-OSM-DE}} 
		& \multicolumn{3}{c}{\texttt{Wikidata-OSM-IT}} 
        	& \multicolumn{3}{c}{\texttt{Average}} \\
\cmidrule(l{0pt}r{6pt}){2-4} \cmidrule(l{0pt}r{6pt}){5-7} \cmidrule(l{0pt}r{6pt}){8-10} \cmidrule(l{0pt}r{0pt}){11-13} 
    & Precision & Recall & F1 
    &  Precision & Recall & F1  
       &  Precision & Recall & F1  
    &  Precision & Recall & F1\\  
\midrule

\textsc{BM25}   	& 45.22 & 42.59 & 43.86
               				 	& 47.28 & 41.60 & 44.26
               				 	& 44.49 & 41.67 & 43.04 
                				& 45.66 & 41.95 &43.72\\
                                
\textsc{Spotlight} &	65.17	&	32.26	&	43.15
                   			 	&	69.79	&	51.03	&	58.95
                    			&	54.79	&	26.89	&	36.08
                        		&	 63.25	&	36.73	&	46.06\\

\textsc{Geo-Dist}   &	74.46	&	74.46	&	74.46 
                    &	62.16	&	62.16	&	62.16 
                    &	72.80	&	72.80	&	72.80
                    & 69.81 	& 69.81 & 69.81\\

\textsc{LGD} & \textbf{100.00} & 44.09 & 61.20 
                & \textbf{100.00} & 47.46 & 64.37 
                & \textbf{100.00} & 43.59 & 60.71
                & \textbf{100.00} & 45.05 & 62.09\\
         
\textsc{LGD-super}      & \textbf{100.00} & 53.25 & 69.50
                        & \textbf{100.00} & 55.34 & 71.25
                        & \textbf{100.00} & 53.79 & 69.95
                        & \textbf{100.00}  & 54.13 & 70.23 \\

\textsc{Yago2Geo}   & 63.66 & 44.98 & 52.71 
                    & 64.48 & 48.61 & 55.43
                    & 58.40 & 47.36 & 52.30
                    & 62.18 & 46.98 & 53.48\\
         
\textsc{Yago2Geo-super} & 78.49 & 47.38 & 59.09
                        & 73.49 & 48.96 & 58.76
                        & 72.25 & 48.73 & 58.20
                        & 74.74 & 48.36 & 58.69 \\

\textsc{LIMES/Wombat}   & 74.03  & 17.50  & 28.31 
                        & 78.54 & 17.01 & 27.97
                        & 65.28 & 17.22  & 27.25
                        & 72.62 & 17.25 & 27.84 \\
                
\midrule

\textsc{Osm2kg-tfidf}   & 95.06 & 90.60 & 92.77
                        & 93.67 & 86.37 & 89.87
                        & 93.98 & 87.07 & 90.39
                        & 94.24 & 88.01 & 91.01 \\

\textsc{Osm2kg}		& 95.51 & \textbf{91.90}  & \textbf{93.67}*
                    & 93.98 & \textbf{88.29} & \textbf{91.05}*
                    & 94.39 & \textbf{88.68} & \textbf{91.45}*
                    & 94.62 & \textbf{89.63} & \textbf{92.05}\\

     \bottomrule
\end{tabular}

    \end{subtable}\vspace{0.2cm}
    \begin{subtable}[c]{\textwidth}
    \caption{Link prediction performance on the DBpedia datasets}
    \label{ev:tab:perfomanceDBpedia}
    \footnotesize
    \begin{tabular}{l@{\quad}rrc@{\hskip 1em}rrc@{\hskip 1em}rrc@{\hskip 1em}rrc}
\toprule
\multirow{2}{*}{\raisebox{-\heavyrulewidth}{\textbf{Approach}}} &               \multicolumn{3}{c}{\texttt{DBpedia-OSM-FR}} 
		& \multicolumn{3}{c}{\texttt{DBpedia-OSM-DE}} 
		& \multicolumn{3}{c}{\texttt{DBpedia-OSM-IT}} 
        	& \multicolumn{3}{c}{\texttt{Average}} \\
\cmidrule(l{0pt}r{6pt}){2-4} \cmidrule(l{0pt}r{6pt}){5-7} \cmidrule(l{0pt}r{6pt}){8-10} \cmidrule(l{0pt}r{0pt}){11-13} 
    & Precision & Recall & F1 
    &  Precision & Recall & F1  
       &  Precision & Recall & F1  
    &  Precision & Recall & F1\\  
\midrule

\textsc{BM25}   	& 70.04 & 69.32 & 69.68
               				 	& 47.28 & 76.84 & 75.58
               				 	& 44.49 & 41.67 & 43.04 
                				& 53.94 & 62.61 & 62.77\\
                                
\textsc{Spotlight} &	72.40	&	49.42	&	58.74
                   			 	&	79.08	&	62.31	&	69.70
                    			&	85.38	&	56.17	&	67.76
                        		&	 78.95	&	55.97	&	65.40\\

\textsc{Geo-Dist}   &	85.94	&	85.94	&	85.94 
                    &	66.49	&	66.49	&	66.49 
                    &	86.17	&	86.17	&	86.17
                    &  79.53	&  79.53 &  79.53\\

\textsc{LGD} & \textbf{100.00} & 61.81 & 76.40
                & \textbf{100.00} & 60.72 & 75.56 
                & \textbf{100.00} & 64.94 & 78.74
                & \textbf{100.00} & 62.49 & 76.90\\

\textsc{LGD-super}      & \textbf{100.00}  & 88.18 & 93.72
                        & \textbf{100.00} & 84.56 & \textbf{91.63}
                        & \textbf{100.00} & 86.90 & 92.99
                        & \textbf{100.00} & 86.55 & 92.78\\

\textsc{Yago2Geo}   & 77.52 & 70.40 & 73.78
                    & 87.41 & 75.84 & 81.22
                    & 94.74 & 78.47 & 85.84 
                    &  86.56 & 74.90 & 80.28 \\

\textsc{Yago2Geo-super} & 84.74 & 82.47  & 83.59
                        & 93.62 & 80.14  & 86.36
                        & 97.46 & 81.28  & 88.64
                        & 91.94 & 81.30 & 86.19 \\

\textsc{LIMES/Wombat}   & 82.34  & 60.33 & 69.64
                        & 79.00 & 68.00 & 73.09
                        & 97.38 & 70.89 & 82.05
                        & 86.24 & 66.41 & 74.93 \\

\midrule

\textsc{Osm2kg-tfidf}   & 98.68 & 95.35 & 96.99
                        & 95.61  & 84.93 & 89.95
                        & 98.46 & 89.83 & 93.95
                        & 97.91 & 90.04 & 93.63\\

\textsc{Osm2kg}   & 99.06 & \textbf{96.25}  & \textbf{97.63}*
                    & 95.65 & \textbf{85.83} & 90.47
                    & 99.11 & \textbf{90.13} & \textbf{94.41}
                    & 97.94 & \textbf{90.74} & \textbf{94.17}\\

     \bottomrule
\end{tabular}

    \end{subtable}
    \label{ev:tab:overAllPerformance}
\end{table*}

The main goal of the evaluation is to assess the link discovery performance of \approach compared to the baselines. Moreover, we analyse the effectiveness of the classification model and the proposed features and perform parameter tuning.

\subsection{Link Discovery Performance}
\label{sec:link_discovery_evaluation}

Table \ref{ev:tab:overAllPerformance} summarises the overall link discovery performance results of the  
\textsc{BM25}, \textsc{SPOTLIGHT}, \textsc{Geo-Dist}, \textsc{LGD}, \textsc{LGD-super}, \textsc{YAGO2GEO}, \textsc{YAGO2GEO-super}, 
and \textsc{LIMES/Wombat} baselines as well as our proposed approach in the \textsc{\approach} and \textsc{\approach-TFIDF} variants. 
Table \ref{ev:tab:perfomanceWikidata} reports the results of the experiments conducted on the link datasets from Wikidata, while Table \ref{ev:tab:perfomanceDBpedia} reports the result on the DBpedia datasets.
We report the macro averages of the 10-fold cross-validation conducted on the corresponding link dataset concerning the precision, recall, and F1 score. 
In our experiments, we observed that the micro averages behave similarly.

Overall, we observe that in terms of F1 score, \approach performs best on all Wikidata datasets, 
where it achieves an F1 score of 92.05\% on average and outperforms the best performing 
\textsc{LGD-super} baseline by 21.82 percentage points.
Furthermore, we observe that \approach achieves the best performance concerning the recall on all datasets.
Moreover, \approach maintains high precision, i.e., 94.62\% on Wikidata and 97.94\% on DBpedia, on average. 
Regarding the DBpedia datasets, we observe that \approach outperforms the baselines on \texttt{DBpedia-OSM-FR} and \texttt{DBpedia-OSM-IT}, whereas the difference to the \textsc{LGD-super} baseline is much smaller, compared to Wikidata. On \texttt{DBpedia-OSM-DE}, \textsc{LGD-super} archives a slightly higher F1 score, compared to \approach. 
This result indicates that, in contrast to Wikidata, the respective DBpedia and OSM datasets are well-aligned in terms of names and geographic coordinates, such that simple heuristics utilising name similarity and geographic distance can already yield good results in many cases.
In contrast, the task of link discovery in Wikidata is more challenging.
In these settings, the advantages of the \approach approach become clearly visible.

The \textsc{BM25} and \textsc{Spotlight} baselines adopt name similarity for matching, whereas \textsc{Spotlight} can also make use of the knowledge graph context, including entity types. 
As we can observe, \textsc{BM25} shows relatively low performance in terms of both precision (on average 45.66\% (Wikidata) and 53.94\% (DBpedia)) and recall (on average 41.95\% (Wikidata) and 62.61\% (DBpedia)). 
The \textsc{Spotlight} baseline can improve on \textsc{BM25} regarding precision and F1 score on Wikidata and DBpedia datasets. However, the absolute precision and F1 scores of \textsc{Spotlight}, with the maximum F1 score of 65.40\% on Wikidata, are not competitive. Overall, we conclude that name similarity, as adopted by these baselines, is not sufficient for effective link prediction. 

\begin{table}
    \centering
    \caption{Parameters learned by the \textsc{LGD-super} and the \textsc{Yago2Geo-super} baselines}
    \footnotesize
\begin{tabular}{lrrrr}
     \toprule
     \multirow{2}{*}{\raisebox{-\heavyrulewidth}{\textbf{Data Set}}} &
     \multicolumn{2}{c}{\textsc{LGD-super}} &
     \multicolumn{2}{c}{\textsc{Yago2Geo-super}}\\
     & $th_{block}$ & $th_{str}$ & $th_{block}$ & $th_{str}$ \\
     
     \midrule
     \texttt{Wikidata-OSM-FR} & 1500 & 0.1 & 1000 & 0.70  \\
     \texttt{Wikidata-OSM-DE} & 2000 & 0.1 & 2000 &  0.80  \\
     \texttt{Wikidata-OSM-IT} & 1500  & 0.1 & 1000 & 0.70  \\
     \texttt{DBpedia-OSM-FR} & 1000 & 0.1 & 1000 & 0.30  \\
     \texttt{DBpedia-OSM-DE} & 5000 & 0.1 & 2000 &  0.75  \\
     \texttt{DBpedia-OSM-IT} & 20000 & 0.3 & 1500 & 0.30 \\
     \bottomrule
\end{tabular}
    \label{tab:superbaselines-parameter}
\end{table}

The \textsc{LGD} and \textsc{LGD-super} baselines that combine name similarity and geographic distance 
achieve the best precision of 100\% on all datasets. 
However, \textsc{LGD} baselines suffer from lower recall.
\textsc{LGD-super} achieves on average 54.13\% recall on Wikidata and 86.55\% recall on DBpedia, 
overall resulting in lower F1 scores on average compared to \approach. 
The \textsc{Yago2Geo} baseline that uses similar features as \textsc{LGD} achieves higher recall scores than \textsc{LGD} (46.98\% on Wikidata, 74.90\% on DBpedia on average) but cannot maintain the high precision of \textsc{LGD} (on average 62.18\% on Wikidata, 86.56\% on DBpedia). 
Overall, \textsc{Yago2Geo} achieves lower F1 scores compared to \approach. 

Regarding the supervised baselines, Table \ref{tab:superbaselines-parameter} presents the parameters learned by \textsc{LGD-super} and the \textsc{Yago2Geo-super} during the training process. 
We observe that \textsc{Yago2\-Geo-super} learns more restrictive parameters, whereas \textsc{LGD-super} allows for less restrictive threshold values. 
This result indicates that the ranking function of \textsc{LGD-super} that combines geographic distance and name similarity is more robust than the ranking function of \textsc{Yago2\-Geo-super}.
\textsc{Yago2Geo-super} uses geographic distance exclusively for blocking and ranks the candidates based solely on the name similarity.
We observe that both baselines achieve a reasonably good performance on the DBpedia datasets. 
On the contrary, both baselines can not reach comparable performance on the Wikidata datasets and result in 70.23\% F1 score for \textsc{LGD-super}, and 58.69\% F1 score for \textsc{Yago2Geo-super}, on average.

\textsc{GEO-DIST}, which solely relies on the geographic distance, achieves an F1 score of 69.81\% on Wikidata, and 79.53\% on DBpedia on average.  
Although a significant fraction of the OSM nodes can be correctly linked solely based on the geographic distance, still a significant fraction of nodes (on average 30.19\% for Wikidata and 20.74\% for DBpedia) can not be appropriately linked this way. We observe that the lower performance of \textsc{Geo-Dist} corresponds to densely populated areas (e.g., large cities), where we expect knowledge graphs to have a higher number of entities, making disambiguation based on geographic distance ineffective.
\approach overcomes this limitation and outperforms the \textsc{GEO-DIST} baseline by 22.24 percentage points (Wikidata) and 14.64 percentage points (DBpedia) on average concerning F1 score.  

The \textsc{LIMES/Wombat} baseline that aims to learn rules for link discovery in a supervised fashion does not achieve competitive performance on any considered dataset and results in 27.84\% F1 score for Wikidata and 74.93\% F1 score for DBpedia on average. 
One of the main reasons for such low performance is that \textsc{LIMES/Wombat} requires all entities to contain all considered properties. As none of the OSM tags is mandatory, this baseline is de-facto limited to only frequently used properties, such as the name and the geo-coordinates. These properties alone are insufficient to extract the rules leading to competitive performance in the link discovery task on these datasets.

Comparing the performance of \approach across the datasets, we observe that scores achieved on the  \texttt{Wikidata\-OSM-FR} and \texttt{DBpedia-OSM-FR} datasets (93.67\%, and 97.63\% F1 score) are higher than on the other language editions. This result can be explained through a more consistent annotation of the nodes within the \texttt{OSM-FR} dataset. For instance, in \texttt{OSM-FR} eight key-value combinations appeared more than 2000 times, whereas in \texttt{OSM-DE} and \texttt{OSM-IT} only two to four combinations are that frequent.

Comparing the overall link discovery performance on the DBpedia and Wikidata datasets, we observe that higher F1 scores are achieved on DBpedia by all considered approaches. 
Furthermore, the \textsc{LGD-super} and \textsc{Yago2Geo-super} baselines that utilise only geographic distance and name similarity heuristics can reach high performance on DBpedia (up to 92.78\% F1 score on average). In contrast, their maximal performance on Wikidata is limited to 70.23\% F1 score.
This result indicates that, in general, geographic coordinates and entity names of OSM are better aligned with DBpedia than with Wikidata. 
This result also suggests that the link discovery task is more difficult on Wikidata. 
Our \approach approach is particularly useful in these settings, where we achieve 21.82 percentage points increase in F1 score compared to the best performing \textsc{LGD-super} baseline.

\subsection{Comparison to OSM2KG-TFIDF}
Comparing the performance of \approach with the \textsc{OSM2kg-tfidf} variant, we observe that the embedding of \approach leads to better performance (1.04 percentage points of F1 score for Wikidata and 0.54 percentage points of F1 score for DBpedia on average).

We observe a statistically significant difference between the F1 scores of \approach and \textsc{Osm2kg-tfidf} on all Wikidata datasets and \textsc{DBpedia-OSM-FR} (paired t-tests with $p < 0.01$).
Through a manual inspection of exemplary instances, we found that \approach especially improves over \textsc{OSM2KG-tfidf} on discovering links for nodes with name information and nodes corresponding to Wikidata types with a small number of instances.
For example, a node corresponding to a private school\footnote{\url{https://www.openstreetmap.org/node/2733503641}} was wrongly assigned to a trade school\footnote{\url{https://www.wikidata.org/wiki/Q828825}} instead of the  entity\footnote{\url{https://www.wikidata.org/wiki/Q2344470}}.
In this example, the name of the OSM node and the geo-entity are identical. We believe that through the high number of dimensions in the TF-IDF representation, the \emph{name} dimension and the corresponding \emph{name similarity} might lose importance, even though the name is typically a very effective feature in the context of link discovery. 
From the \textsc{Random Forest} models, we observe that the \emph{name similarity} achieves a lower mean decrease impurity \cite{10.5555/2999611.2999660}  in \textsc{Osm2kg-tfidf} than in \approach, indicating the lower contribution of the feature.
Moreover, the \emph{KV-embedding} poses a distributed representation of the OpenStreetMap tags. We believe that especially for Wikidata types with a small number of instances the distributed representation might be more robust, whereas in a TF-IDF representation single tags could introduce bias towards types with a higher number of instances. 
In the example above, the tag \texttt{toilets:wheelchair=yes} is likely to co-occur with both the private school and trade school types but might be biased towards the more populated type.

We do not observe statistically significant differences between 
\textsc{OSM2KG} and \textsc{OSM2KG-tfidf}
on the \linebreak \texttt{DBpedia-OSM-DE} and \texttt{DBpedia-OSM-IT} datasets.
On these datasets, baselines that exclusively make use of geographic distance and name similarity such as \textsc{LGD-super} achieve the best or close-to-best F1 score.
Therefore, the individual importance of the \emph{KV-embedding} or the TF-IDF feature is not as high as for the other datasets.

\begin{figure}
    \centering
    \includegraphics[width=0.48\textwidth]{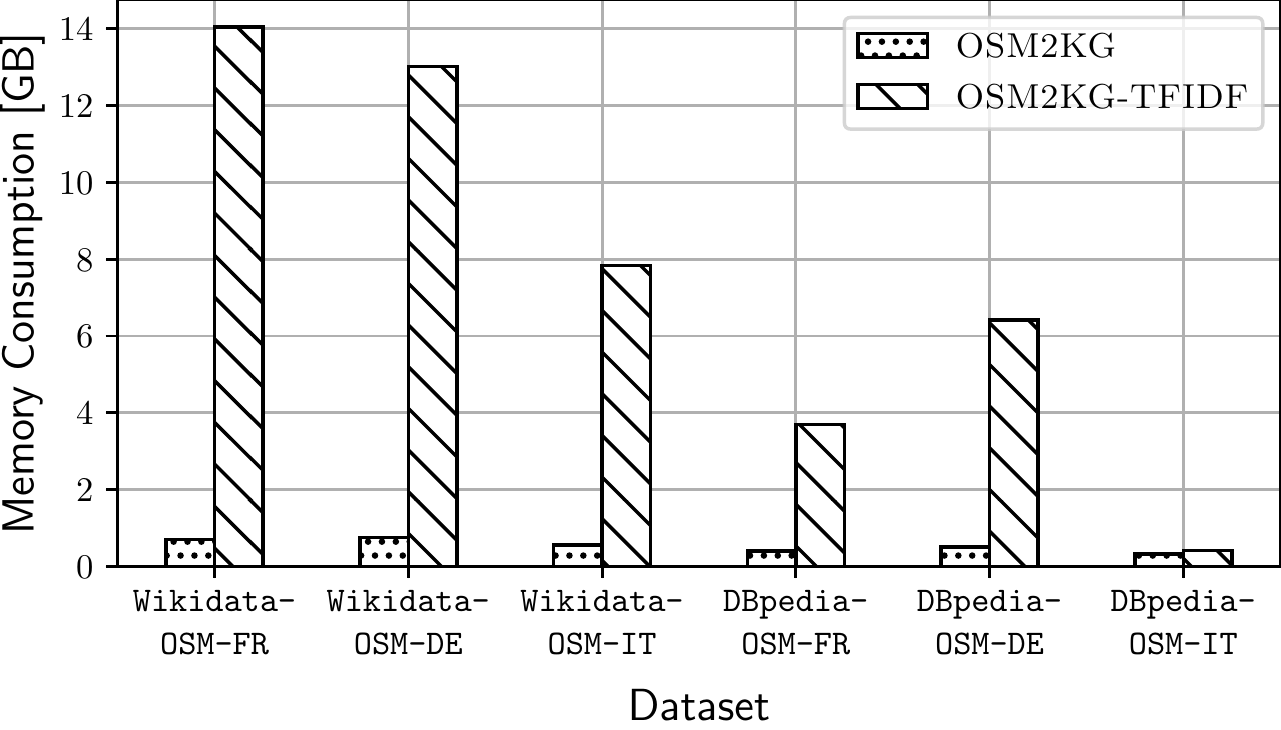}
    \caption{Average memory consumption across folds of the training of the \textsc{Random Forest} models used by \approach and \textsc{OSM2kg-tfidf}.}
    \label{fig:memory_consumption}
\end{figure}

Furthermore, the proposed \emph{KV-embedding} provides a compact representation that consists of only 3-5 dimensions, whereas the corresponding TF-IDF representations consist of 1000 dimensions.
Figure \ref{fig:memory_consumption} contrasts the average memory consumption across the folds of the random forest models of \approach and \textsc{OSM2kg-tfidf}.
We observe that the usage of the \emph{KV-embedding} generally results in a lower memory footprint than the TF-IDF variant, which becomes particularly visible for larger datasets.
The difference is largest on the \texttt{Wikidata-OSM-FR} dataset, where the \emph{KV-embedding} (0.7 GB) requires only 5\% of memory compared to the TF-IDF variant (14 GB).
We observe the smallest difference on \texttt{DBpedia-OSM-IT}. This dataset has the smallest number of instances (2353), resulting in the small memory difference between the models (0.1 GB).

We conclude that KV-embedding is an effective, concise, and task-independent way to represent the OSM information. 
We believe that this representation makes OSM data more usable for models that may suffer from the curse of dimensionality or memory limitations.

\begin{table*}
    \centering
    \caption{Comparison of \approach F1 scores [\%] with respect to the classification model, best scores are bold.}
\footnotesize
\centering
\begin{tabular}{lrrrr||rrrr}
\toprule
\textbf{Classifier} & \makecell{\texttt{Wikidata-}\\\texttt{OSM-FR}} & \makecell{\texttt{Wikidata-}\\\texttt{OSM-DE}} & \makecell{\texttt{Wikidata-}\\\texttt{OSM-IT}} &  \makecell{\texttt{Wikidata-}\\\texttt{Average}} &
 \makecell{\texttt{DBpedia-}\\\texttt{OSM-FR}} & \makecell{\texttt{DBpedia-}\\\texttt{OSM-DE}} & \makecell{\texttt{DBpedia-}\\\texttt{OSM-IT}} & \makecell{\texttt{DBpedia-}\\\texttt{Average}}  \\
\midrule
\textsc{Random Forest} & 93.67 & 91.05 & \textbf{91.45} & 92.05 
    & \textbf{97.63} & \textbf{90.47} & 94.41 & \textbf{94.17} \\
\textsc{Decision Tree} & \textbf{94.45} & \textbf{91.17} & 91.01 & \textbf{92.21}
    & 97.12 & 89.62 & \textbf{94.56} & 93.77\\
\textsc{Na\"ive Bayes} & 70.88   &  63.64 & 66.45 & 66.99
    & 76.69   & 77.69 & 88.40 & 80.93\\
\textsc{Logistic Regression} & 65.36 & 66.40 & 70.87 & 67.54
    & 86.84 & 86.93 & 88.71 & 87.49\\
\bottomrule
\end{tabular}

    \label{ev:tab:classifierComp}
\end{table*}

\begin{table*}
    \centering
    \footnotesize
    \caption{Differences in \approach F1 score [percentage points] when leaving out single features using \textsc{Random Forest}.}
    \footnotesize
\begin{tabular}{lrrrr||rrrr}
     \toprule
     \textbf{Left out Feature} & \makecell{\texttt{Wikidata-}\\\texttt{OSM-FR}} & \makecell{\texttt{Wikidata-}\\\texttt{OSM-DE}} & \makecell{\texttt{Wikidata-}\\\texttt{OSM-IT}} & \makecell{\texttt{Wikidata-}\\\texttt{Average}} &
      \makecell{\texttt{DBpedia-}\\\texttt{OSM-FR}} & \makecell{\texttt{DBpedia-}\\\texttt{OSM-DE}} & \makecell{\texttt{DBpedia-}\\\texttt{OSM-IT}} & \makecell{\texttt{DBpedia-}\\\texttt{Average}}  \\
     \midrule
     \text{KV-embedding} & 2.80  & 3.91 & 
     4.53 & 3.75
        & 1.94 & 1.96 & 0 & 1.30
     \\
     \text{Geo Distance} & 15.28 & 14.72 & 11.98 & 13.99 
        & 2.81 & 2.19 & 8.67 & 4.56\\
    Name & 1.92 & 3.52 & 3.51 & 2.98 
        & 3.61 & 5.66 & 6.86 & 5.38\\
     \text{Entity Type} & 0.71  & 2.00 &  2.77 & 1.83 
        & 0.45 & 0.54 & -0.08 & 0.30\\
     \text{Popularity} &  0.29 & 1.07 & 0.94 & 0.77
        &  0.29 &  0.28 & -0.02 & 0.18 \\
     \text{Entity Type} \& \text{Popularity} & 1.67 & 9.30 &  6.94 & 5.97 
        & 0.84 & 1.50 & -0.08 & 0.75\\ 
     \bottomrule
\end{tabular}

    \label{ev:tab:featureEval}
\end{table*}

\subsection{Classification Model Performance}
\label{sec:classifier_evaluation}

Table \ref{ev:tab:classifierComp} presents the F1 scores achieved by \approach with respect to each dataset while varying the classification model. 
In particular, we evaluate the performance of \textsc{Random Forest}, \textsc{Decision Tree}, \textsc{Na\"ive Bayes}, and \textsc{Logistic Regression}. 
As we can observe, the performance of the classification models is consistent among the datasets. 
\textsc{Random Forest} and \textsc{Decision Tree} achieve similar F1 scores and show the best performance, i.e., on average 92.05\% (Wikipedia), 94.17\% (DBpedia) F1 score using \textsc{Random Forest}, 
and 92.21\% (Wikidata), 93.77\% (DBpedia) using \textsc{Decision Tree}.
According to a paired t-test, the observed differences between the \textsc{Random Forest} and \textsc{Decision Tree} are not statistically significant on our datasets.
In contrast, the performance of \textsc{Na\"ive Bayes} and \textsc{Logistic Regression} is much lower, i.e., they achieve on average only 66.99\% (Wikidata), 80.93\% (DBpedia) F1 score using \textsc{Na\"ive Bayes} and 67.54\% (Wikidata), 87.49\% (DBpedia) using \textsc{Logistic Regression}.

We conclude that non-linear classification models such as \textsc{Random Forest} and \textsc{Decision Tree} are better suited to the problem we address than the linear models.
This result also suggests that the classification problem is not linearly separable.
In our experiments in Section \ref{sec:link_discovery_evaluation}, we made use of \textsc{Random Forest} classification models.

\subsection{Feature Evaluation}
\label{sec:feature_evaluation}

In this section, we assess the feature contributions of \approach. 
To assess the contribution of the single features to link discovery, we conducted a leave-one-out feature evaluation. In particular, we removed each feature individually from the feature set and determined the difference in F1 score to quantify the feature importance.

Table \ref{ev:tab:featureEval} shows the differences in the F1 score of the \approach model when a single feature is left out compared to the F1 score achieved when the entire feature set is used. 
Since no difference is negative,  except for \texttt{DBpedia-OSM-IT}, we conclude that all features typically contribute to better classification performance.
\textit{Geo Distance} results in the most substantial difference of 13.99 percentage points on average for Wikidata.
On DBpedia, \textit{Geo Distance} results in the second-largest difference of 4.56 percentage points on average. 
The most considerable difference for DBpedia results from the \emph{Name} feature, with 5.38 percentage points on average. For Wikidata, the \emph{Name} feature results in a variation of 2.98 percentage points on average. 
The importance of the \emph{Name} feature on DBpedia indicates that the names of the OSM and DBpedia datasets are well-aligned.
This result confirms our observations in Section \ref{sec:link_discovery_evaluation}, 
where we discussed the performance of the \textsc{LGD-super} baseline that utilises both features.
The \textit{KV-embedding} feature shows the second-largest difference on Wikidata (3.75 percentage points) and the third-largest difference on DBpedia (1.30 percentage points) on average.
As expected, the contribution of this feature is higher for the more complex link discovery task in Wikidata, as opposed to DBpedia, where simple heuristics may suffice. 
As an extreme example, we do not observe any contribution of \textit{KV-embedding} for \texttt{DBpedia-OSM-IT}.
As discussed before, simple heuristics (e.g., geographic distance and name similarity) are sufficient to achieve relatively high performance on this dataset.
The \textit{Entity Type} and \textit{Popularity} show the smallest differences, where \textit{Entity Type} has slightly larger differences than \textit{Popularity}. 
For the Wikidata datasets, we observe that the individual contributions of the features are rather small, i.e. 1.83 percentage points (\textit{Entity Type}) and 0.77 percentage points (\textit{Popularity}) on average.
When leaving both features out, we observe a difference of 5.97 percentage points on average. We conclude that the information encoded in both features is partly redundant. Furthermore, this relatively large difference indicates feature importance.
We conclude that for Wikidata datasets the information of the \textit{Entity Type} is especially useful when combined with the \textit{Popularity} feature.
On the contrary, for the DBpedia datasets, we observe that the contribution of the \textit{Popularity} feature is nearly identical to the joint contribution of \textit{Entity Type} and \textit{Popularity}. 
For \texttt{DBpedia-OSM-IT} we observe negative contributions for both features. 
Again, this indicates that geographic distance and name similarity are sufficient for link discovery in this dataset.
Although \textit{Entity Type} and \textit{Popularity} are correlated in many cases, they can provide complementary information for some instances. Intuitively, the joint information can help to disambiguate entities similar concerning one of the features, but dissimilar regarding the other. For example, two railway stations of different sizes are likely to be described with a different number of statements, whereas the type is identical. In such cases, in addition to the Entity Type, Popularity can help to disambiguate entities better.

\subsection{Parameter Tuning}
\label{sec:parameter_evaluation}

We evaluate the influence of the parameters such as embedding size and the blocking threshold value on the performance of \approach.

\subsubsection{Embedding Size}
\label{sec:embedding_size_evaluation}

\begin{figure}
    \centering
    \includegraphics[width=0.48\textwidth]{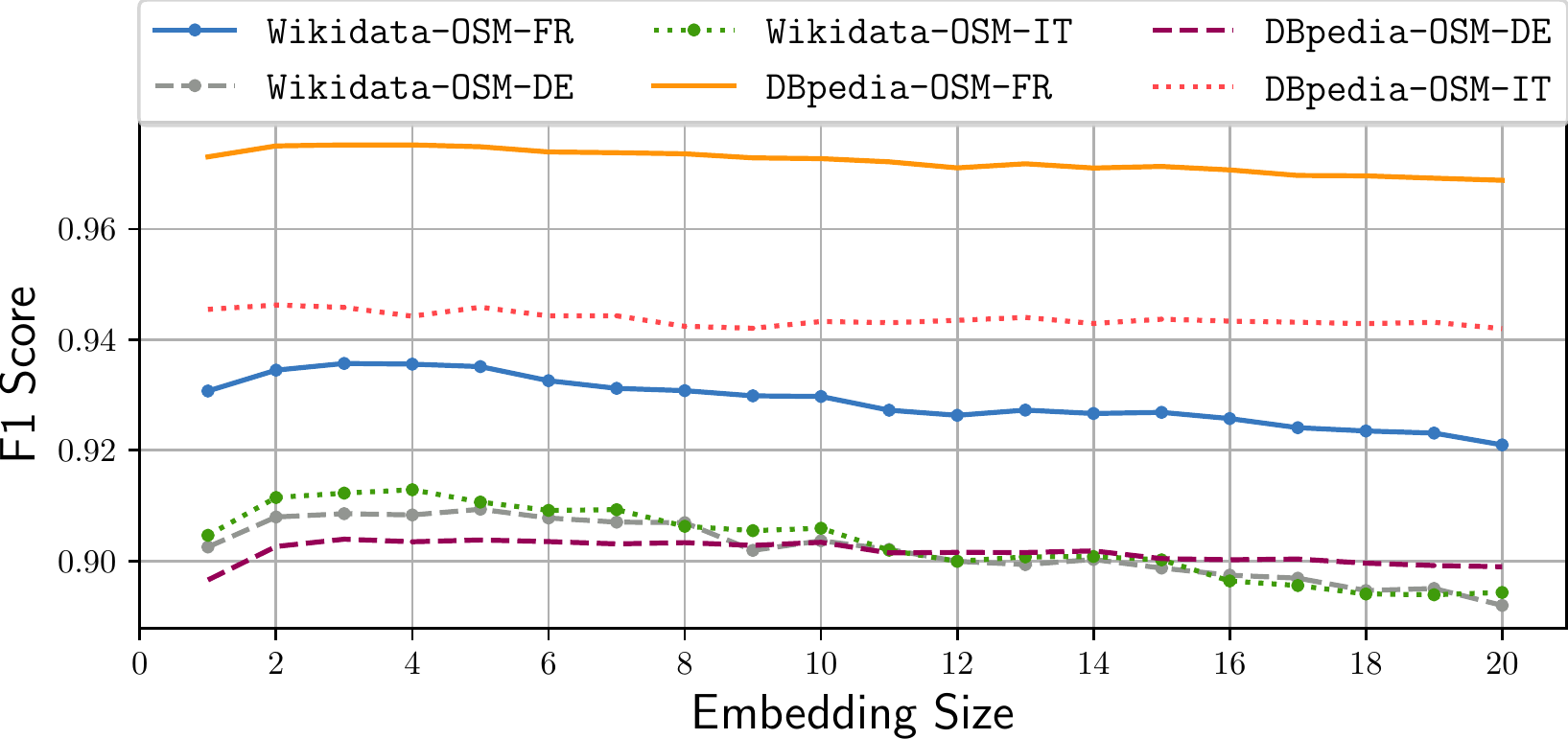}
    \caption{Influence of the embedding size on F1 score of the \textsc{Random Forest} classifier.}
    \label{ev:fig:EmbSize}
\end{figure}

\begin{figure*}
    \centering
    \includegraphics[width=\textwidth]{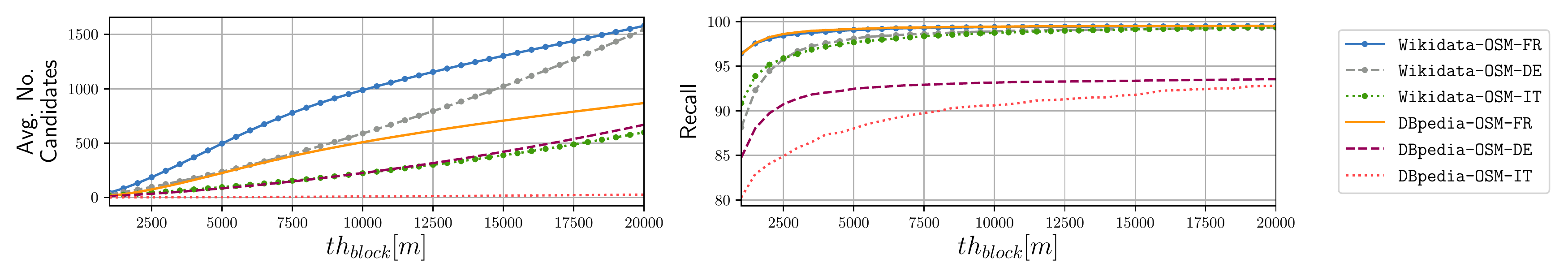}
    \caption{Influence of the threshold $th_{block}$ on the average number of candidates and recall of the blocking step.}
    \label{ev:fig:thblock}
\end{figure*}

The embedding size corresponds to the number of dimensions (i.e. neurons) in the projection layer of the neural model presented in Section \ref{sec:approach:key-value-embedding}.
Figure \ref{ev:fig:EmbSize} shows F1 scores obtained with respect to the number of dimensions of the \emph{KV-embedding} achieved by the \textsc{Random Forest} classifier on all datasets.

We observe similar trends for all datasets except for \texttt{DBpedia-OSM-IT}.
Overall, we can observe a growth of the F1 score of the classifier with an increasing number of dimensions, between one and four dimensions for all datasets.
We conclude that embeddings with an insufficient number of dimensions are not able to capture all relevant information. When the number of dimensions increases, more information can be encoded, which leads to better performance. 
As we can observe, the curve achieves its maximum at three dimensions for the \texttt{Wikidata-OSM-FR}, and \texttt{DBpedia-OSM-FR} datasets, at four dimensions for \texttt{Wikidata\-OSM-IT} and at five dimensions for the \texttt{Wikidata-OSM-DE} and \texttt{DBpe\-dia-OSM-DE} datasets.
Further increase of the embedding size does not lead to an increase in performance. On the contrary, the performance can drop, indicating that no additional beneficial information is obtained by adding further dimensions. 

For \texttt{DBpedia-OSM-IT}, we observe a near-constant performance around 94\% F1 score of the classifier. 
As discussed in Section \ref{sec:feature_evaluation}, here the contribution of the KV-embedding is not as high as for the other datasets. Thus the variation of the embedding size does not result in any significant performance changes for this dataset.

Overall, we conclude that 3-5 dimensions are most suited for the datasets that make effective use of the KV-embedding feature. Thus we adopted the following number of dimensions: \texttt{Wikidata-OSM-FR}: 3, \texttt{Wikidata-OSM-DE}:5, \texttt{Wikidata\\-OSM-IT}: 4, \texttt{DBpedia-OSM-FR}: 3, \texttt{DBpedia-OSM-DE}: 5, \linebreak \texttt{DBpedia-OSM-IT}: 4. 

\subsubsection{Blocking Threshold}
\label{sec:blocking_threshold_evaluation}

The blocking threshold $th_{block}$ represents the maximal geographic distance considered for candidate entity generation, as discussed in Section \ref{sec:candidate_entity_generation}. For a single OSM node, all knowledge graph entities that are closer than $th_{block}$ are considered as candidates. 
The value of $th_{block}$ can be determined experimentally by evaluating the recall of the blocking step.

Figure \ref{ev:fig:thblock} shows the influence of  $th_{block}$ on the average number of candidates and the recall of the blocking step. 
Considering the average number of candidates, we observe a linear-like rise (i.e., the slope of the curve is nearly constant) of the number of candidates concerning $th_{block}$ for all datasets, whereas the datasets differ in slope.
Due to the low geographic density of the \texttt{DBpedia-OSM-IT} dataset, the corresponding slope is especially low.
Concerning recall, we observe that the curve starts with a steady incline, but quickly saturates with an increasing $th_{block}$. We conclude that in most cases, the correct candidate exhibits a geographic distance of about 2.5 km. 
Thus, in our experiments, we chose $th_{block}=2.5$ km.
This threshold value allows for more than 85\% recall of correct candidates for the DBpedia datasets and 95\% recall for the Wikidata datasets in the blocking step, while effectively limiting the number of candidates.
For \texttt{DBpedia-OSM-IT}, we adopt a different $th_{block}$ threshold of 20 km to increase recall on this dataset.

\begin{figure*}
    \centering
    \includegraphics[width=0.7\textwidth]{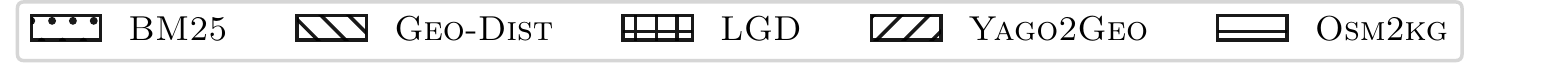}
    \begin{subfigure}{0.3\textwidth}
        \centering
        \includegraphics[width=\textwidth]{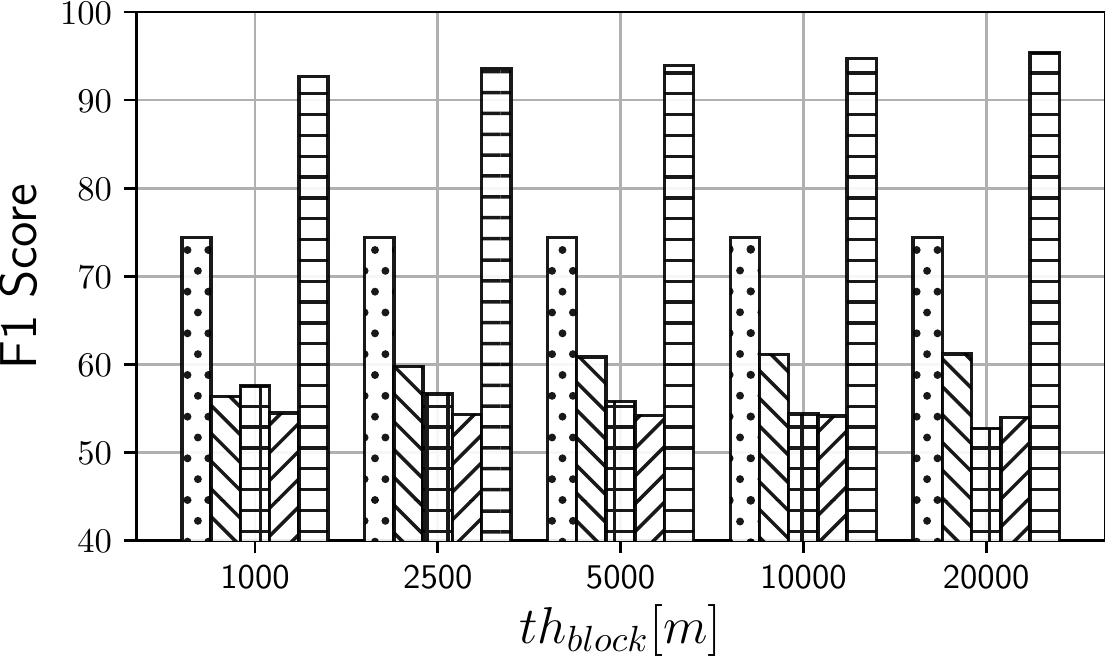}
        \caption{\texttt{Wikidata-OSM-FR}}
    \end{subfigure}
    \begin{subfigure}{0.3\textwidth}
        \centering
        \includegraphics[width=\textwidth]{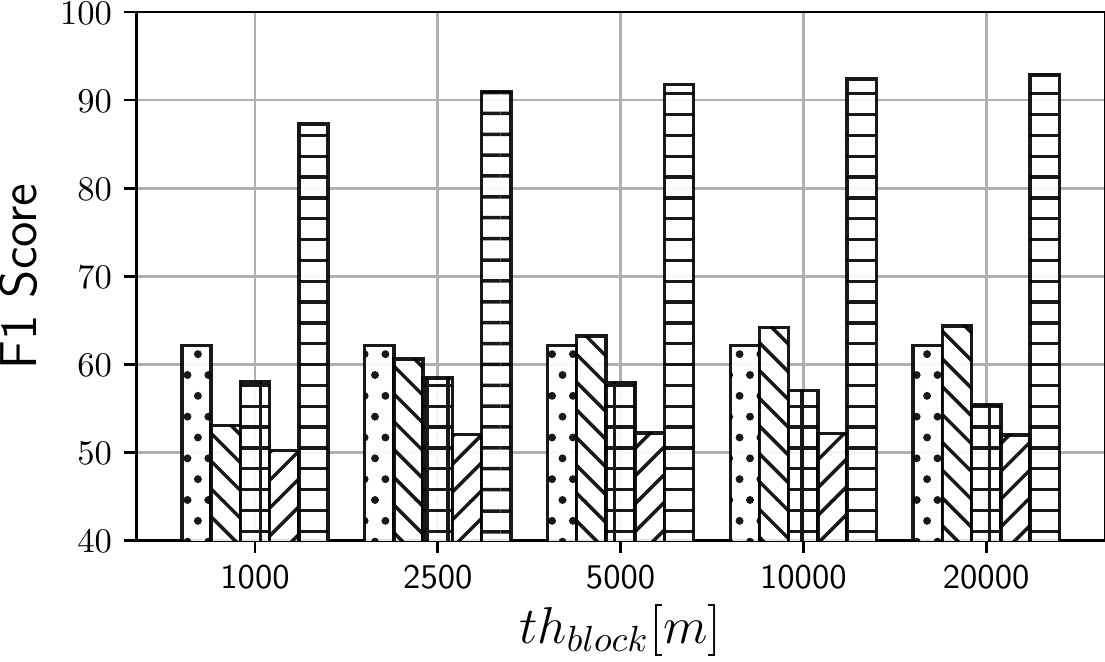}
        \caption{\texttt{Wikidata-OSM-DE}}
        \label{fig:compare_block:w-de}
    \end{subfigure}
    \begin{subfigure}{0.3\textwidth}
        \centering
        \includegraphics[width=\textwidth]{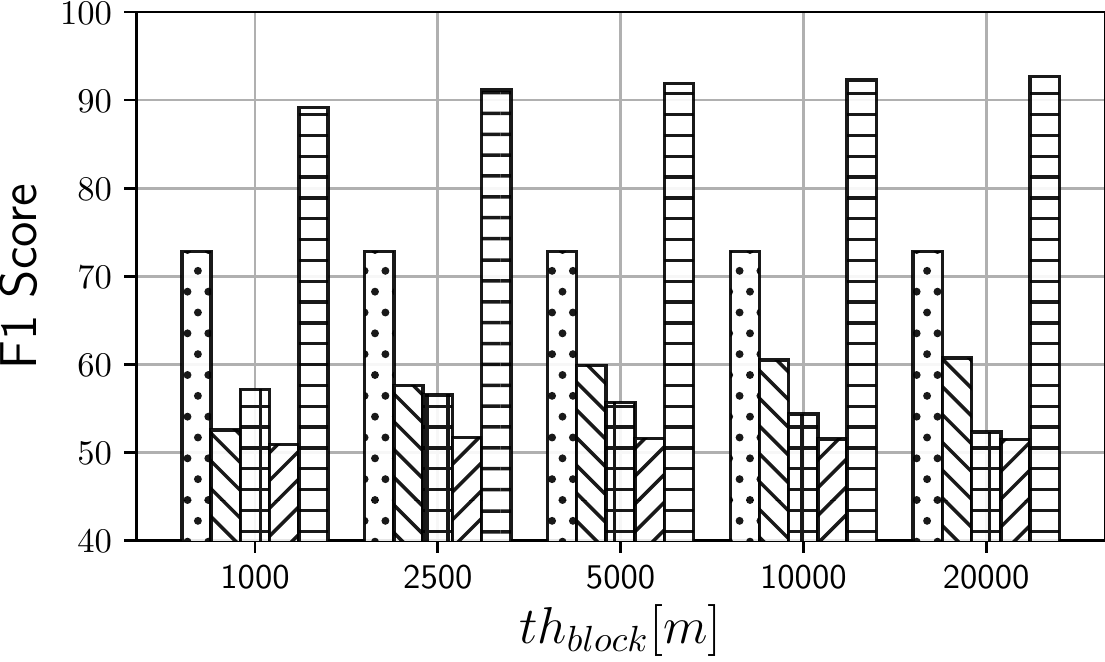}
        \caption{\texttt{Wikidata-OSM-IT}}
    \end{subfigure}
    \vspace{0.2cm}

    \begin{subfigure}{0.3\textwidth}
        \centering
        \includegraphics[width=\textwidth]{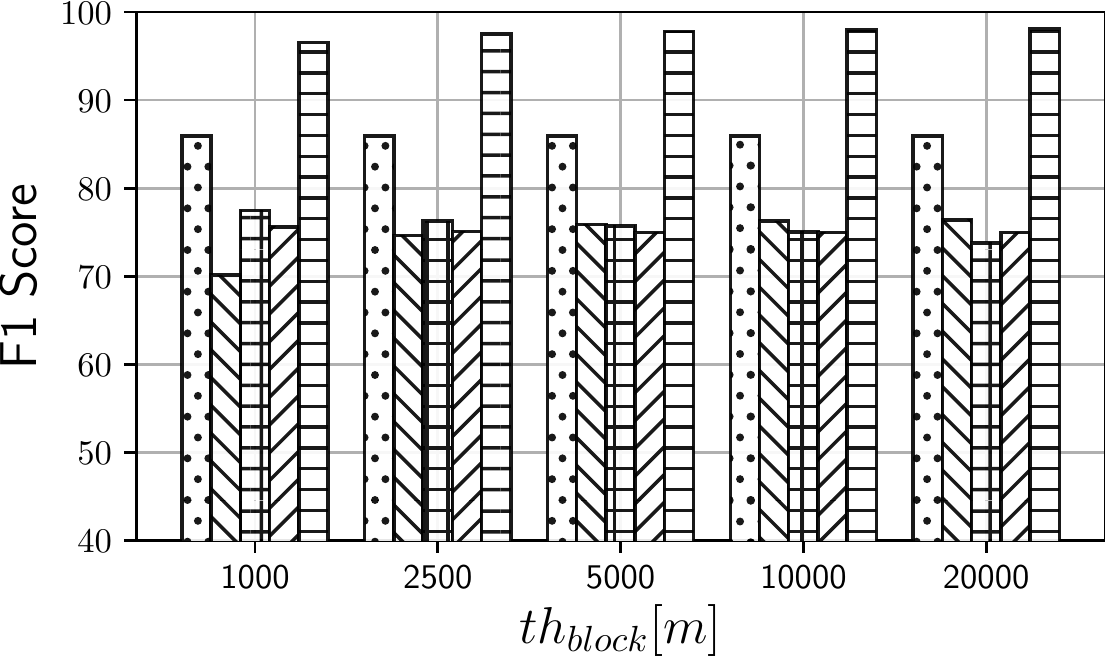}
        \caption{\texttt{DBpedia-OSM-FR}}
    \end{subfigure}
    \begin{subfigure}{0.3\textwidth}
        \centering
        \includegraphics[width=\textwidth]{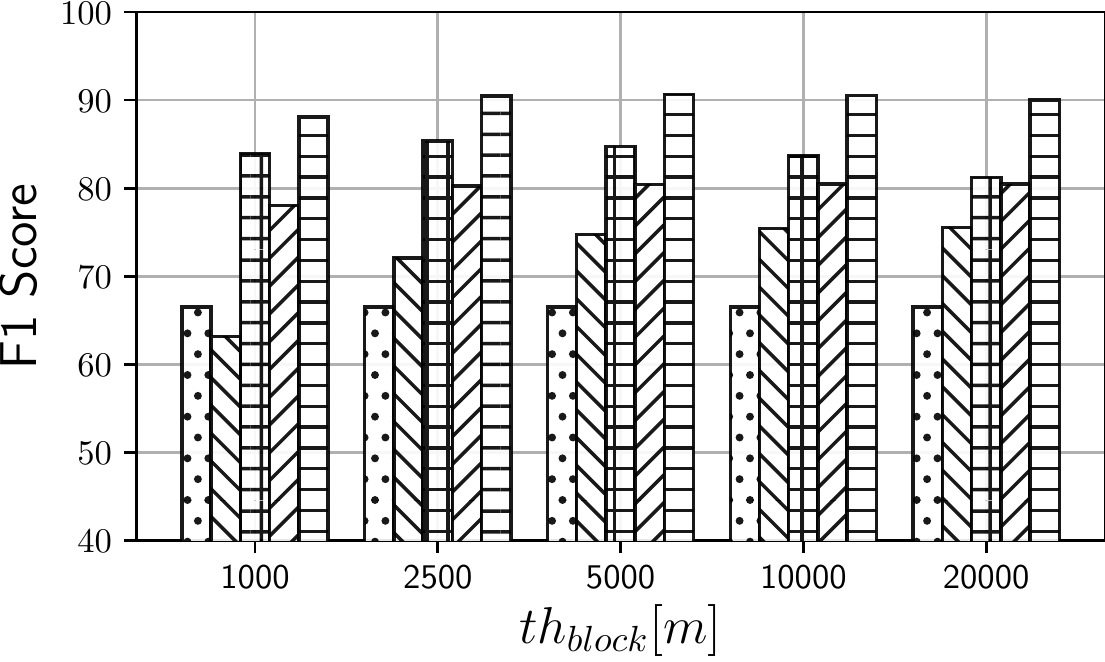}
        \caption{\texttt{DBpedia-OSM-DE}}
        \label{fig:compare_block:d-de}
    \end{subfigure}
    \begin{subfigure}{0.3\textwidth}
        \centering
        \includegraphics[width=\textwidth]{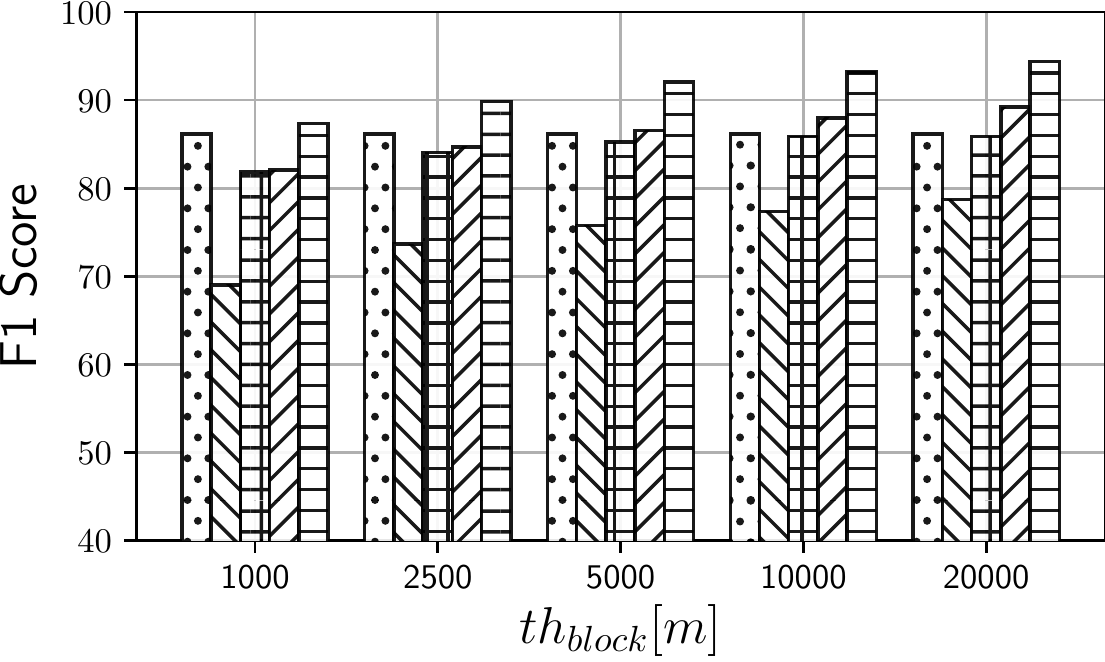}
        \caption{\texttt{DBpedia-OSM-IT}}
        \label{fig:compare_block:d-it}
    \end{subfigure}
    \caption{Link discovery performance concerning $th_{block}$ value for \approach and the baselines that can include a blocking step. X-axis presents the value of $th_{block}$ in meter. Y-axis presents the F1 score.}
    \label{fig:compare_block}
\end{figure*}

To make the impact of geospatial blocking comparable across the considered approaches, we assess the effect of the blocking step on the overall link discovery performance. 
To this extent, we added an additional blocking step to the \textsc{BM25} and \textsc{Geo-Dist} baselines and evaluate the models \textsc{BM25}, \textsc{Geo-Dist}, \textsc{LGD}, \textsc{Yageo2Geo} and \approach with the blocking thresholds $th_{block} \in $ \{1, 2.5, 5, 10, 20\} km.
Figure \ref{fig:compare_block} presents the F1 scores regarding the blocking threshold value $th_{block}$.
As we can observe, the general link discovery performance is not very sensitive to the $th_{block}$ value.
However, if $th_{block}$ value is chosen too low, e.g. 1 km, the link discovery performance can drop, as shown in Figure \ref{fig:compare_block:w-de}.
Overall, an optimal threshold value depends on the model as well as on the dataset.  
For example, \textsc{LGD} may benefit from a lower blocking threshold value, as shown in Figure \ref{fig:compare_block:d-de}, whereas \textsc{Geo-Dist} works better with a higher threshold (Figure \ref{fig:compare_block:d-it}).
For \approach we do not observe any significant impact for values of $th_{block} \geq 2.5$ km for most datasets.
For the supervised variants of the baselines \textsc{LGD} and \textsc{Yago2Geo},  \textsc{LGD-super} and \textsc{Yago2Geo-super}, we observe that the appropriate threshold can be determined during the training process.
The performance of the \textsc{Geo-Dist} baseline is degraded with the limitation of the additional blocking step, as this limitation does not contribute to precision, but potentially limits recall of this baseline.
The BM25 baseline benefits from the blocking step but is still clearly outperformed by \approach.
In summary, as presented by Figure \ref{fig:compare_block}, we observe that \approach outperforms all baselines for all values of the blocking threshold $th_{block}$ on all considered datasets concerning F1 score.

\subsection{Error Analysis}

\begin{table}
\centering
\caption{Distribution of error types on nodes for which no correct link could be found by \approach. 
}
    \resizebox{.48\textwidth}{!}{
\begin{tabular}{lrrrrr}
     \toprule
     \textbf{Error Type} & \makecell{\texttt{Wikidata-}\\\texttt{OSM-FR}} & \makecell{\texttt{Wikidata-}\\\texttt{OSM-DE}} & \makecell{\texttt{Wikidata-}\\\texttt{OSM-IT}} &
     \texttt{Avg.} \\
     \midrule
         No candidate found & 41\% & 54\% & 54\% & 49.67\% \\
     \makecell[l]{Wrong candidate selected} & 39\% & 37\% & 22\% & 32.67\% \\
     Duplicate entity in Wikidata & 17\% & 4\% & 20\% & 13.67 \% \\
     Wrong link in ground truth & 3\% & 5\% & 4\% & 4.00\% \\
    \bottomrule
    \end{tabular}
}

    \label{tab:error}
\end{table}

We conducted an error analysis through manual inspection of a random sample of 100 nodes for which \approach identified no correct link for each of the Wikidata datasets.
Table \ref{tab:error} presents the resulting error distribution.
As we can observe, the most common reason for errors is a too restrictive candidate selection leading to an empty candidate set (in 49.67\% of cases), 
followed by the selection of wrong candidates (in 32.67\% of cases) and quality issues in Wikidata such as duplicate entities (in 13.67\%) as well as wrong links in the ground truth data (in 4\%).
Note that the restrictive candidate selection is subject to the choice of the blocking threshold value. 
For this study, the threshold was chosen in such a way that 95\% recall of the blocking step was achieved.
In a small number of cases (3\% on average), the candidate set is not empty, but the correct candidate is not included in this set.
This issue can be addressed by an adaptive increase of the threshold for the nodes without any candidates.

Furthermore, we observe that the selection of wrong candidates in most cases happens within the regions with a high geographic density of Wikidata entities, e.g., in cities where single houses can represent entities, resulting in a large candidate set. 
To further increase the precision of \approach, a dedicated, supervised model for geographically dense regions can be trained. Such a model can follow a more restrictive policy, e.g., by requiring higher confidence to establish a link.

Finally, the detection of duplicate entities and wrong ground truth links indicates the potential to adopt \approach for de-duplication of geo-entities in Wikidata to increase data quality.
These observations provide a basis for an incremental tuning of \approach in future work.

\subsection{Discussion}
\label{sec:discussion}

Approaches that mainly rely on name similarity heuristics and do not leverage any geospatial features are not suitable for effective link prediction for the OSM nodes. We can observe this by considering the relatively low performance of the \textsc{BM25} and \textsc{SPOTLIGHT} baselines, where  \textsc{SPOTLIGHT} achieved F1 scores of 46.06\% (Wikidata) and 65.40\% (DBpedia), on average. 
Geospatial features such as geographic distance are a reliable indicator to match OSM nodes with knowledge graph entities in our datasets. This observation is confirmed by the \textsc{Geo-Dist} baseline, which reached F1 scores of 69.81\% (Wikidata) and 79.53\% (DBpedia) by solely considering the geographic distance. However, in a significant fraction of cases, geospatial information alone is insufficient to disambiguate OSM nodes effectively.
Heuristics using a combination of the name similarity and geospatial features, 
and in particular the supervised \textsc{LGD-super} baseline, can achieve competitive performance on the DBpedia datasets.
However, they are insufficient for link discovery in more complex datasets, such as Wikidata, where the entity names are not well-aligned with OSM. 

The proposed \approach approach combines the latent representation of OSM nodes that captures the semantic similarity of the nodes with geospatial information and is highly effective for link prediction. \approach is of particular advantage for link discovery between OSM and Wikidata, where it significantly outperforms the baselines concerning the recall and F1 score.  
Overall, we observe that the proposed latent node representation as \emph{key-value embedding} combined with geospatial distance is an effective way to facilitate link discovery in a schema-agnostic volunteered geographic dataset such as OSM. This representation, with only 3-5 dimensions, is compact and task-independent. 

Limitations in link discovery can arise from the candidate generation step, where we consider the set of entities for which geographic coordinates are available in the knowledge graph. A promising direction for future research is to discover identity links between OSM nodes and geographic entities for which geographic coordinates are not available in the knowledge graph.

In this work, we focused the discussion and evaluation of \approach on Wikidata and DBpedia as target knowledge graphs due to their openness, popularity, and availability of training data (i.e., the links between these knowledge graphs and OSM). Nevertheless, the proposed \approach approach is applicable to other knowledge graphs provided a set of identity links between OSM and the target knowledge graph is available for training the \approach classifier.

\section{Related Work}
\label{sec:relatedWork}

\textit{Link Discovery} is the task of identifying semantically equivalent resources in different data sources \cite{DBLP:journals/semweb/NentwigHNR17}. 
Nentwig et al. \cite{DBLP:journals/semweb/NentwigHNR17} provide a recent survey of link discovery frameworks, with prominent examples, including Silk \cite{www2009227} and LIMES \cite{Ngomo:2011:LTA:2283696.2283783}. 

In particular, the Wombat algorithm, integrated within the LIMES framework \cite{10.1007/978-3-319-58068-5_7}, is a state-of-the-art approach for link discovery in knowledge graphs.
Link discovery approaches that operate on Linked Data typically expect datasets in 
Resource Description Framework (RDF) format having a schema defined by an underlying ontology and data exhibiting graph structure. This assumption does not apply to the OSM data represented as key-value pairs. 

Besides the syntactic and structural differences, LIMES relies on several assumptions that severely limit its applicability to OSM datasets.
First, LIMES assumes a one-to-one mapping between properties. 
In contrast, the required mappings between the Wikidata properties and the OSM keys are 1:n, as a Wikidata property can correspond to several OSM keys. For example, the “instanceOf” property in Wikidata corresponds to “place,” “natural,” “historic,” and many other keys in OSM. Second, LIMES requires all instances to contain all considered properties. Therefore LIMES is limited to utilise only frequently used properties, such as the name and the geo-coordinates. To this end, LIMES is not suited to utilise the information from other infrequent properties for mapping. Finally, the current LIMES implementation does not adequately support a combination of different data types, such as strings and geo-coordinates. Given these differences, the application of LIMES to the OSM data is de-facto restricted to the name matching.
We utilise Wombat/LIMES as a baseline for the evaluation. 
Our experimental results confirm that \approach outperforms this baseline.

In the context of individual projects such as LinkedGeoData and Yago2Geo \cite{SLHA11, 10.1007/978-3-030-30796-7_12}, a partial transformation of OSM data to RDF was conducted using manually defined schema mappings for selected keys. In contrast, the proposed \approach approach adopts an automatically generated latent representation of OSM data.

\textit{Entity linking} (also referred to as entity disambiguation) is the task of linking mentions of real-world entities in unstructured sources (e.g., text documents) to equivalent entities in a knowledge base. 
A recent survey on entity linking approaches is provided in \cite{6823700}.
Entity linking approaches typically adopt Natural Language Processing (NLP) techniques and use the context of the entity mentions such as phrases or sentences. However, such a context is not available in OSM, where textual information is mainly limited to node labels (typically available as a specialised name tag).
One of the most popular state-of-the-art models to automatically annotate mentions of DBpedia entities in natural language text is \emph{DBpedia Spotlight} \cite{Daiber:2013:IEA:2506182.2506198}. DBpedia Spotlight adopts NLP techniques to extract named entities (including locations) from text and uses a context-aware model to determine the corresponding DBpedia entities. This approach serves as a baseline in our experiments, whereas we use the name tag of an OSM node as its textual representation. 

\textit{Linking geographic data:}
The most relevant projects in the context of our work are  LinkedGeoData \cite{SLHA11} and Yago2Geo \cite{10.1007/978-3-030-30796-7_12}.
LinkedGeoData is an effort to lift OSM data into semantic infrastructure. This goal is addressed through deriving a lightweight ontology from the OSM tags and transforming OSM data to the RDF data model. LinkedGeoData interlinks OSM nodes represented as RDF with geo-entities in external knowledge sources such as DBpedia and GeoNames. 
Yago2Geo aims at extending the knowledge graph YAGO2 \cite{HOFFART2013} with geographic knowledge from external data sources. To this extent, identity links between YAGO2 and OSM are computed. 
Both interlinking approaches rely on manually defined schema mappings and heuristics based on name similarity and geographic distance.  
The dependence of both approaches on manual schema mappings restricts the coverage of mapped entity types and can also negatively affect link maintenance. 
In contrast, the \approach approach proposed in this article 
extracts latent representations of OSM nodes fully automatically. 
The LinkedGeoData and Yago2Geo interlinking approaches serve as a baseline in our experiments. 
Our experimental results confirm that \approach outperforms both baselines.
The applications of linked geographic data include, for example, the training of comprehensive ranking models \cite{DESSI2016366} or the creation of linked data based gazetteers \cite{CARDOSO2016389}.

\emph{Geospatial link discovery} \cite{DBLP:conf/semweb/SavetaFFN17, AHMED2018139, SherifDSN17, SmerosK16} refers to the problem of creating topological relations across geographic datasets. These links express the topographic relations between entities (e.g., intersects and overlaps). 
For example, \cite{SmerosK16} presented the problem of discovery of spatial and temporal links in RDF datasets. In Radon \cite{SherifDSN17}, efficient computation of topological relations between geospatial resources in the datasets published according to the Linked Data principles was presented.
In contrast, in this work, we focus on link discovery for identity links.

\textit{Geographic representation learning:} Recently, several approaches emerged that employ representation learning to encode geographic data. Typical data sources are point of interest and floating car data, where the proposed architectures include graph embeddings \cite{Xie:2016:LGP:2983323.2983711, Wang:2017:RRL:3132847.3133006, Yang:2019:RUM:3308558.3313635}, metric embeddings \cite{LIU2018183},
stacked autoencoders \cite{Ma:2018:PRE:3269206.3271733}, generative models \cite{HUANG20201119}, and word2vec-like models \cite{AAAI1714902, 10.1145/3308558.3313456}. 
\cite{KejriwalS17} proposed neural embeddings for Geonames that explicitly takes the geospatial proximity into account.
The proposed \approach approach relies on an embedding architecture inspired by word2vec to automatically encode semantic similarity of the OSM nodes using key-value pairs.
The embedding aims to generate a similar representation for the nodes with similar properties, independent of their location. Thus, we do not include location information in the embedding.

\section{Conclusion}
\label{sec:conclusion}

In this article, we proposed \approach, a novel link discovery approach to predict identity links between OpenStreetMap nodes and geographic entities in knowledge graphs. \approach combines latent representations of heterogeneous OSM nodes with a supervised classification model to predict identity links across large-scale, diverse datasets effectively. 
Our experiments conducted on three large-scale OSM datasets for Germany, France, and Italy and Wikidata and DBpedia knowledge graphs demonstrate that the proposed \approach approach can reliably discover identity links. 

\approach achieves an F1 score of 92.05\% on Wikidata and of 94.17\% on DBpedia on average, which corresponds to a 21.82 percentage points increase in F1 score on Wikidata compared to the best performing baselines.

Whereas we conducted our evaluation on OSM, Wikidata and DBpedia, our approach can be applied to other VGI sources and knowledge graphs as long as a training set of identity links is available. 
In future work, we would like to develop novel applications that take advantage of integrated geographic and semantic information created by \approach.
Furthermore, we would like to explore the applicability of the proposed \emph{KV-embedding} to further datasets and tasks.

\subsubsection*{Acknowledgement} 
 This work is partially funded by the DFG, German Research Foundation (``WorldKG", DE 2299/2-1, 424985896), the Federal Ministry of Education and Research (BMBF), Germany (``Simple-ML", 01IS18054), (``Data4UrbanMobility'', 02K15A040), and the Federal Ministry for Economic Affairs and Energy (BMWi), Germany (``d-E-mand", 01ME19009B).

\bibliographystyle{elsarticle-num}
\bibliography{references}

\end{document}